\documentclass[nofootinbib,aps,prd,showpacs]{revtex4-1}

\usepackage{hyperref}
\hypersetup{
	colorlinks=true,
	linkcolor=blue,
	filecolor=red,      
	citecolor=BurntOrange,
	urlcolor=Blue,
}
\usepackage{amssymb}        
\usepackage{amsmath}        
\usepackage{graphicx}
\usepackage{color}        
\usepackage{float}
\usepackage[x11names,dvipsnames]{xcolor}
\usepackage{caption}
\usepackage{subcaption}

\def\nn{\nonumber}

\begin{document}

	\title{ Aschenbach effect for spinning particles in Kerr-(A)dS spacetime }
	
	\author{Ali Vahedi}
	\affiliation{
	Department of Astronomy and High Energy Physics,\\ Faculty of Physics, Kharazmi University, P.~O.~Box 15614, Tehran, Iran.}
	\email{vahedi@khu.ac.ir}
	
	\author{Jafar Khodagholizadeh}%
	\affiliation{Farhangian University, P.O. Box 11876-13311, Tehran, Iran%
	}%
	\email{gholizadeh@ipm.ir}

	\author{Arman Tursunov}
	
	\affiliation{%
		Research Centre for Theoretical Physics and Astrophysics, Institute of Physics, Silesian University in Opava, Bezru{\v c}ovo n{\'a}m.13, CZ-74601 Opava, Czech Republic
	}%
	\affiliation{Institute of Experimental and Theoretical Physics, Al-Farabi Kazakh National University, Almaty 050040, Kazakhstan}
	
	\email{arman.tursunov@physics.slu.cz}

	\date{\today}
	
	\begin{abstract}
		A non-monotonic behavior of the velocity gradient of a test particle revolving around a rapidly rotating black hole in the locally non-rotating frame of reference is known as the Aschenbach effect. This effect can serve as a distinguishing signature of rapidly rotating black holes, being potentially useful for the measurements of the astrophysical black hole spins. This paper is the generalization of our previous research to the motion of spinning particles around a rotating black hole with non-zero cosmological constant. We show that both the particle's spin $s$ and the cosmological constant $\Lambda$ modify the critical value of the black hole spin $a_c$, for which the Aschenbach effect can be observed; $a_c$ can increase or decrease depending on the signs of $s$ and $\Lambda$. We also found that the particle's spin $s$ can mimic the effect of the cosmological constant $\Lambda$ for a given $a_c$, causing thus a discrepancy in the measurements of $s$, $\Lambda$ and $a_c$ in the Aschenbach effect. 
	\end{abstract}

\maketitle

\section{Introduction}

Currently available experimental data related to the multiwavelength observations of both stellar mass and supermassive black hole candidates suggest no convincing deviation of the black hole spacetime metric from that of the rotating Kerr spacetime \cite{Abbott:2016blz,Eckart:2017bhq}.
On the other hand, the cosmological observations indicate that our universe at the present state is undergoing accelerated expansion. One of the most plausible cosmological models that fit the observational data is the so-called $\Lambda$-CDM model \cite{Condon:2018eqx} describing the universe with the cosmological constant $\Lambda$ filled with cold dark matter (CDM). Therefore, at the cosmological scales, this gives a rise to the generalization of the spacetime metric of astrophysical black holes from Kerr to Kerr-de-Sitter (Kerr-dS) spacetime. It is also useful to explore the cases with an opposite sign of the cosmological constant, namely the Kerr-anti-de-Sitter, motivated by the famous anti-de Sitter/conformal field theory (AdS/CFT) correspondence \cite{Guica:2008mu}. Thus, in the case of Kerr-(A)dS spacetime the metric is described solely by three parameters: mass  $M$ and spin $a$ of the black hole and the cosmological constant $\Lambda$.

The most fundamental among the parameters of a black hole is its mass, which in many cases is measured through the direct observations of nearby objects with relatively high accuracy. For example, the mass of a supermassive black hole located at the center of the Milky Way have been measured by near-infrared observations of S-stars cluster with the closest S0-2 star moving around central object Sgr~A* with the consistent mass of the order of $4 \times 10^{6} M_{\odot}$ \cite{Abuter:2018drb}.  

In contrast to the mass, the measurement of the black hole spin is a more difficult task, as the gravitational effect of the spin has no Newtonian analog. Therefore, the spin measurements require the observations of effects occurring relatively close to the black hole event horizon, i.e. in a strong gravity regime.  Current methods of the spin estimations are model-dependent giving different values in various models.    Among the promising avenues for the spin determination of astrophysical black hole is potential gravitational waves detections from extreme mass ratio inspirals (EMRI) by future space-based Laser Interferometer Space Antenna (LISA) \cite{Babak:2017tow}. Great progress in the determination of black hole spins has been achieved in recent years using the X-ray spectroscopy by modeling and fitting the spectra of black hole accretion disks \cite{McClintock:2011zq}. The estimates of spins of black holes obtained by such methods suggest that a large number of black holes have very high spin values, especially those located in the active galactic nuclei (AGN) \cite{Laura}. Moreover, in many cases, the observed luminosities are explained by the presence of near-extremal or extremal Kerr black holes. From the theoretical point of view rapidly rotating black holes with near-extremal or extremal spins are also subjects of special attention in connection to so-called Kerr/CFT correspondence \cite{Guica:2008mu} with near-horizon enhanced symmetry, giving interesting predictions. Therefore, 
it is especially important to seek for the potentially observable effects occurring in the vicinity of near-extremal rotating black holes. 

One of such effects is the Aschenbach effect \cite{Aschenbach:2004kj,Aschenbach:2006cj}, which we describe below.
In the field of a rotating Kerr black hole, the velocity of a test particle  $v(r;a)$ moving in the co-rotating circular geodesics measured in the locally non-rotating frame (LNRF) has a monotonous radial profile satisfying $\partial v/\partial r < 0$, if the spin parameter of the black hole is $a<0.9953$. However, when $a>0.9953$ the behavior of the velocity $v(r;a)$ close to the black hole horizon becomes non-monotonic, i.e. gradient of velocity changes its sign in the near-horizon region. This effect has been discovered by B. Aschenbach in mid 2000s  \cite{Aschenbach:2004kj,Aschenbach:2006cj} and named after him.  
This effect is potentially observable when high-resolution observations of matter revolving in the immediate vicinity of the event horizon of supermassive black holes become available. In this sense, the direct observations of the flares and dynamics of flare components, similar to those, recently discovered at the Galactic center by GRAVITY@ESO \cite{2018A&A...618L..10G} seems promising \cite{2020ApJ...897...99T}. 
If the spin of a supermassive black hole is near-extremal (with $a>0.9953$), the Aschenbach effect can be observed for co-rotating matter through the corresponding change of the radiation flux at different radial positions of accretion disk or the flares. Another interesting possibility to observe the Aschenbach effect is through the detection of gravitational waves from EMRI by future LISA experiment \cite{Babak:2017tow}.  Note, that the Aschenbach's spin limit for Kerr black hole is still below the realistic maximum of $a_{\rm BH} = 0.998$ obtained by \cite{1974ApJ...191..507T} as the equipartition of the gravitational and magnetic energy in the black hole accretion disk.

%
 For a test particle moving around rapidly rotating Kerr black hole with the spin $a>0.9953$, Aschenbach in \cite{Aschenbach:2006cj} have also shown that the orbital radius below $1.8$ of gravitational radii, where the rate of change of the LNRF velocity has a hump, nearly coincides with the location, where the radial and vertical epicyclic frequencies have the radio close to $3:1$. Such a coincidence can be directly related to the observational phenomenon of the high-frequency quasi-periodic oscillations (QPOs) \cite{Stuchlik:2007sta} of X-ray flux observed in many systems containing compact objects, including black holes. It is widely believed that the QPOs phenomena observed with twin peaks are originated in the non-linear resonant epicyclic oscillations occurring around black holes or neutron stars \cite{Stuchlik:2006xa,Kluzniak:2001abr,Abramowicz:2003bbk,Torok:2005aks,Kolos:2017ojf,Tursunov:2018H}.



The non-monotonic behavior of the orbital velocity 
has been  widely studied in the literature, including e.g.
a non-geodesic motion with constant angular momentum  \cite{Stuchlik:2004wk}, braneworld generalizations of the Kerr spacetime \cite{Stuchlik:2011sw}, motion of charged particle in magnetized black holes \cite{Tursunov:2016dss,2020Univ....6...26S} among others.  Recently, we have shown that a similar effect occurs also in the case of a spinning particle moving around rapidly rotating Kerr black hole \cite{Khodagholizadeh:2020sex}.

In the test particle approach, a neutral particle following geodesics is considered as a point particle with the mass and size negligible in comparison with those of the central black hole. 
However, one can observe situations, in which the test particle can be interpreted as an extended body, like a star, hot-spot, neutron star, or even a stellar mass black hole revolving around a supermassive black hole in a binary system with extreme mass ratio. In such cases, the test bodies can be spinning, whose spin interaction with the gravitational field of the central black hole leads to the deviation of their trajectories from geodesics. The motion of the particle, in this case, is governed by the Mathisson-Papapetrou-Dixon equations \cite{Mathisson1937,Papapetrou1951,Dixon1964}. Usually, the higher-order spin contributions can be neglected, leaving only the spin-dipole contribution. Such consideration is known as the “pole-dipole” approximation, where
the pole stands for the particle's mass and a dipole for the internal angular momentum, i.e. spin of the particle. 

In this paper we study the dynamics of a spinning particle in the vicinity of a rotating black hole with non-vanishing cosmological constant, focusing on (but not restricted to) the generalization of the Aschenbach effect in the Kerr-(A)dS metric spacetime. For the motion of spinning particle, we use the Mathisson-Papapetrou-Dixon equations with some supplementary restrictions on the spin within the “pole-dipole” approximation. We show that 
this effect survives in $ \Lambda $ fluid which is described by Kerr-de Sitter or Kerr-anti-de Sitter solutions \cite{Mueller:2007tf,Slany:2007gp}. It seems an interplay between the cosmological constant $ \Lambda $ and the black hole spin controls the depth of the " maximum-minimum structure". Note that the range of  Kerr-(A) de Sitter black hole parameters is extended to $ a>1 $ for which the humpy behavior exists.\\
Also by considering the spin of a test particle and using the Mathisson-Papapetrou-Dixon equations for a massive spinning particle the Aschenbach effect have been investigated \cite{Khodagholizadeh:2020sex}. In addition to the black-hole spin, the absolute value of the particle’s spin also plays an important role in the Aschenbach effect. In this case the "minimum-maximum" structure appears only for  $ -0.11 < S/(M \mu) < 1.05 $. In this work, we extend our previous work from the Kerr background to the Kerr-(A)de Sitter metric by correcting the linear relation between the angular momentum of the black hole, particle's spin, and the cosmological constant.

 This work is organized as follows: In section II, we review the MPD equations with supplementary conditions.  In III we investigate the spinning particles in equatorial plane in Kerr-(A) dS background and obtain the angular velocity in zero angular momentum observer's frame (ZAMO). In  section IV the Aschenbach effect 
for three cases: zero spinning, spinning particle with zero-$\Lambda$ and spinning probe with non-vanishing $\Lambda$ background is classified and relation between particle spin, $\Lambda$ and critical angular momentum of black hole is obtained based on the zero, first and second derivatives of the angular velocity conditions. We also discuss the position of the innermost stable circular orbit (ISCO) in dependence on the corresponding parameters. In section V we give concluding remarks and finalize the paper with relevant appendices. 

\section{Review on Mathisson-Papapetrou-Dixon equations}

The motion of a spinning probe particle in the space-time with Riemann curvature tensor $ R^{\mu}{}_{\nu \sigma \tau}$ would be determined by the following three
Mathisson-Papapetrou-Dixon (MPD) equations\footnote{We consider MPD equations at pole-dipole approximation.}~\cite{Mathisson1937,Papapetrou1951,Dixon1964} :
\begin{subequations}\label{MPD}
	\begin{equation}
		\dfrac{D}{d\tau} x^{\mu}(\tau) = u^{\mu}(\tau) \, ,\label{eq:MPD1}
	\end{equation}
	\begin{equation}
		\dfrac{D}{d\tau} p^{\mu}(\tau) = - \dfrac{1}{2} R^{\mu}{}_{\nu \sigma \tau}
		u^{\nu} S^{\sigma \tau} \, ,\label{eq:MPD2}
	\end{equation}
	\begin{equation}
		\dfrac{D}{d\tau} S^{\mu \nu} = p^{\mu} u^{\nu}-p^{\nu} u^{\mu} \, ,\label{eq:MPD3}
	\end{equation}
\end{subequations}
where $\tau$ is a proper time, $x^\mu(\tau)$ is a worldline of the probe particle,	$p^\mu$ is the 4-momentum and $S^{\mu \nu}$ is an antisymmetric spin tensor of the probe particle.  The 4-velocity of a test particle $u^{\alpha}$ is normalized by the condition 
\begin{equation}
	g_{\mu \nu} u^{\mu} u^{\nu} = 1 \, .
	\label{eq:proper}
\end{equation}
We also define the rest mass of particle as
\begin{equation}
	m := u^{\rho} p_{\rho} \, , 
	\label{eq:mmu}
\end{equation}
which is no longer a constant of the motion \citep[see details in][]{Semerak:1999qc}. Therefore, from \eqref{eq:MPD3} we have
\begin{align}\label{eq:mom}
	p^\mu= m u^\mu& + u_\nu \dfrac{D}{d\tau} S^{\mu \nu}\nn\\
	=m u ^\mu&-\frac{S^{\mu\,\nu}R_{\nu \alpha\beta\gamma}p^\alpha  S^{\beta\,\gamma}}{2(\mu^2+\frac{1}{4}R_{\tau \rho\sigma\lambda}S^{\tau \rho}S^{\sigma\lambda})}
\end{align}
where $\mu ^2 = p^{\mu}p_{\mu}$. This momentum-velocity relation means that in general for spinning particle the momentum $p^\mu$ is not parallel to $u^\mu$. An extra term in \eqref{eq:mom}, namely, $u_\nu \dfrac{D}{d\tau} S^{\mu \nu}$, is a  hidden momentum contribution due to the particle's spin. 
For determining the solutions of the MPD equations in a given background geometry, we need supplementary or boundary conditions. 
The most famous among the boundary conditions are Tulczyjew-Dixon (TD) \cite{Tulczyjew1959,Dixon1970} and Frankel-Pirani(FP) \cite{Frenkel1926,Pirani1956} conditions. The FP condition reads 
\begin{equation}\label{con:FP}
	u_{\mu} S^{\mu \nu} = 0 \, ,
\end{equation} 
while the TD condition is given by particle's momentum
\begin{equation}\label{con:TD}
	p_{\mu} S^{\mu \nu} = 0 \, .
\end{equation} 
With TD condition for MPD, one gets two constants of motion:
\begin{subequations}
	\begin{equation}\label{eq:mu}
		\mu ^2 = p^{\mu}p_{\mu} \, ,
	\end{equation}
	\begin{equation}\label{eq:S}
		S^2 = - \dfrac{1}{2} \, S^{\mu \nu} S_{\mu \nu} \, ,
	\end{equation}
\end{subequations}
where $\mu$ is interpreted as the mass of the particle in the center-of-momentum system. Using these conditions one can define the spin vector through the following equation
\begin{equation}\label{eq:spinvec}
	S^{\rho \tau} = \dfrac{1}{\mu} 
	\varepsilon ^{\rho \tau \sigma \nu} p_{\sigma} S_{\nu}, 
\end{equation}
hence the TD constrain, given by \eqref{con:TD} can be reduced to
\begin{equation}\label{con:nTD}
	S_{\rho} p^{\rho} = 0 \, .
\end{equation}
In the following section we discuss the properties of the MPD equations with the TD condition in the spacetime of the Kerr-(A)dS black hole. The solution of equations with FP condition in general would be different but a linear order terms with respect to the Moller radius as a result would be the same \citep[see details in][]{Chicone:2005jj}.

\section{Spinning particles  in the equatorial plane of Kerr-(A)dS  spacetime }
We consider a spinning particle as a probe in the  Kerr-(A)dS  metric  as a background in the standard Boyer-Lindquist coordinates \cite{CARTER1968}, 
\begin{equation}\label{metricall}
	g_{\mu \nu} dx^{\mu} dx^{\nu}
	=\frac{1}{\rho^2 \chi^2}\left[-\Delta_r \left(dt-a \sin^2\theta d\varphi\right)^2+\Delta_\theta \sin^2\theta  \left(a dt-(a^2+r^2) d\varphi\right)^2 \right]+\rho^2\left[\frac{dr^2}{\Delta_r}+\frac{  d\theta^2}{\Delta_\theta}\right],
\end{equation} 
where
\begin{align}
	\rho^2&=r^2+a^2 \cos^2 \theta\nonumber\\
	\chi^2&=1+\frac{\Lambda a^2}{3}\nonumber\\
	\Delta_r&=(r^2+a^2)(1-\frac{\Lambda r^2}{3})-2 M r\nonumber\\
	\Delta_\theta&=\, (1+\frac{\Lambda a^2 \cos^2\theta }{3}). 
\end{align}
Here $\Lambda$ is the cosmological constant and $\Lambda=0$ the metric  \eqref{metricall} coincides with the metric of the Kerr black hole of the mass $M$ and rotational parameter $a$. We restrict the motion to the equatorial plane, with a spin vector perpendicular to this plane, so we have
\begin{equation}\label{eq:eqat}
	\theta = \pi /2 
	\, , \quad
	u^{\theta} = 0 
	\, . \quad
\end{equation}
From (\ref{eq:S}), \eqref{eq:spinvec} and \eqref{con:nTD}
we get 
\begin{equation}\label{eq:Stheta}
	S^r = 0 
	\, , \quad
	S^t = 0 
	\, , \quad
	S^{\varphi} = 0 
	\,\quad S^{\theta} =- \dfrac{S}{r}, 
\end{equation}
which from MPD equations we get $p^\theta=0$. So the non-vanishing components of the spin tensor are\footnote{See also \cite{Zhang:2018omr}.}
\begin{equation}\label{eq:spinvector}
	S^{tr}=-S^{rt}=\dfrac{\chi^2 S p_{\varphi}}{\mu r} 
	\, , \quad
	S^{\varphi t}=- S^{t \varphi}=\dfrac{\chi^2\,S p_r}{\mu r} 
	\, , \quad
	S^{r \varphi}= - S^{\varphi r}= \dfrac{\chi^2\,S p_t}{\mu r} \, .
\end{equation} 
These are in agreement with \cite{Zhang:2018omr}, for $\Lambda=0$, Eq.(\ref{eq:spinvector}) coincides with the results obtained in \cite{Hackmann:2014tga}.  
Further we contract Eq.(\ref{eq:MPD2}) with  $\xi_a$ and Eq.\eqref{eq:MPD3} with $\frac{1}{2}\nabla_a\xi_b$. Summing two equations we get 
\begin{equation}
	\frac{D}{ds}\left( p^a \xi_a +\frac{1}{2} S^{ab} \nabla_a\xi_b\right)=0\,,
\end{equation}
if $\xi$ is considered as a Killing vector \citep[see, e.g.][for more details]{Ehlers1977,Semerak:1999qc,p2015}. So, in addition to constants $\mu$ and $S$, we have extra constants of motion:
\begin{equation}
	E_\xi=p_a \xi^a+\frac{1}{2}S^{ab}\nabla_a \xi_a\,.
\end{equation}
With two Killing vectors, corresponding to the time translation and axisymmetry of the line element \eqref{metricall}, we get two extra constants of motion, namely the energy and angular momentum 
\begin{subequations}
	\begin{equation}\label{eq:E}
		E = p_t- \dfrac{MS}{\mu r^3} \Big(  a p_t + p _{\varphi} \Big) \Big(1-\frac{\Lambda r^3}{3 M} \Big) \, ,
	\end{equation}
	\begin{equation}\label{eq:J}
		J = -p_{\varphi} + \dfrac{S}{\mu} \Big( p_t - \dfrac{aM}{r^3}(1-\frac{\Lambda r^3}{3 M} )
		\big(  p _{\varphi} + a p_t \big) \Big) \, .
	\end{equation}
\end{subequations}
Without loss of generality, we redefine our constants as
\begin{equation}
	\tilde{S}= \dfrac{S}{\mu} \, , \quad
	\tilde{E}= \dfrac{E}{\mu} \, , \quad
	\tilde{J}= \dfrac{J}{\mu} - \Big( \dfrac{S}{\mu} + a \Big) \dfrac{E}{\mu} ,
\end{equation}
and also we define
\begin{equation}\label{effctiv:mass}
	\tilde{M}(r)= (1-\frac{\Lambda r^3}{3 M}) M 
\end{equation}
The last equation is an effective mass corresponding to the gravitational Newton's formula corrected with a cosmological constant $F=\frac{M G}{r^2}-\frac{c^2 \Lambda\, r}{3}=\frac{\tilde{M}(r)G}{r^2}$, \citep[see, e.g.][]{Kerr:2003bp}. 
With \eqref{effctiv:mass}, the equations \eqref{eq:E} and \eqref{eq:J} reduce to the energy and angular momentum in the Kerr space-time \cite{Khodagholizadeh:2020sex}.  Then, using (\ref{eq:E}), (\ref{eq:J}) and (\ref{eq:mu}),  three non-zero 
components of the four-momentum, namely, $p_t$, $p_{\varphi}$ and $p_r$, can be expressed in terms of the constants of motion and the radius $r$ as follows 
\begin{equation}\label{eq:pt}
	\dfrac{p_t}{\mu} = \tilde{E} - \dfrac{\tilde{M}(r) \tilde{S} \tilde{J}}{r^3-\tilde{M}(r) \tilde{S}{}^2}
	\, ,
\end{equation} 
\begin{equation}\label{eq:pphi}
	\dfrac{p_{\varphi}}{\mu} = - \tilde{J} \,
	\dfrac{\big( r^3-a\tilde{M}(r) \tilde{S} \big)}{\big( r^3-\tilde{M}(r) \tilde{S}{}^2 \big)}
	- a  \tilde{E}
	\, ,
\end{equation} 
\begin{equation}\label{eq:pr}
	\dfrac{ \Delta_{r}^2   p_r^2}{\chi^2\,r^4 \mu ^2} =
	\tilde{E} {}^2 -2 \tilde{E} \tilde{J} v(r) - \tilde{J} {}^2 w(r) - u(r)
\end{equation} 
where
\begin{equation}\label{eq:u}
	u(r) =
	\dfrac{\Delta_r }{\chi^2 r^2}
	\, ,
\end{equation} 
\begin{equation}\label{eq:v}
	v(r) =
	\dfrac{ar+\tilde{M}(r) \tilde{S}}{r^3-\tilde{M}(r) \tilde{S}{}^2}
	\, ,
\end{equation} 
\begin{equation}\label{eq:w}
	w(r) =
	\dfrac{r^2 \Delta_r-\big(a r + \tilde{M}(r)\tilde{S}\big)^2}{\big( r^3-\tilde{M}(r) \tilde{S}{}^2 \big) ^2}
	\, .
\end{equation} 
Considering a circular motion, i.e., taking $u^r=0$, from MPD equations we find that $p^r=0$\footnote{See Eq.\eqref{v-p-kerrds} in the Appendix.} and $S^{\varphi t}=0$. Finally from  Eq.\eqref{eq:MPD3} we have
%
\begin{equation}
	- S u ^{\varphi} p ^{\varphi}   + \dfrac{\tilde{M}(r) S}{r^3} \big( u^t-a u^{\varphi} \big) 
	\big( p^t - a p^{\varphi} \big) = 
	\big( p^t - a p^{\varphi} \big) u^{\varphi}  - p ^{\varphi} \big(u^t-a u^{\varphi} \big)
	\, .
	\label{eq:newEq1}
\end{equation}
Inserting $p^t $ and $p^\varphi$ from Eqs.\eqref{eq:pt} and \eqref{eq:pphi} into  Eq.\eqref{eq:newEq1} we find the generalized Keplerian angular velocity\footnote{Note the difference between the upper and lower indecies.} $\Omega=\dfrac{u^\varphi}{u^t}$:
\begin{equation}\label{eq:Omega0}
	\Omega = \dfrac{j \, w(r)+ v(r)}{1 - j \, v(r)
		+(a + \tilde{S} ) \big( j \, w(r)+ v(r) \big) }
	\, .
\end{equation}  
where we have defined $j=\dfrac{\tilde{J}}{\tilde{E}}$. In order to find constant angular momentum $J$ and energy $E$ in terms of the radius $r$ for the motion at the equatorial plane, we solve equations $p_r=0$ and derivative of the right hand side of Eq.\eqref{eq:pr}, which gives 
\begin{equation}\label{eq:circ1}
	0 =
	\dfrac{\tilde{E} {}^2}{w(r)} -2 \tilde{E} \tilde{J} \dfrac{v(r)}{w(r)}
	- \tilde{J} {}^2 - \dfrac{u(r)}{w(r)} \, ,
\end{equation} 
\begin{equation}\label{eq:circ2}
	0 =
	\tilde{E} {}^2 w'(r) 
	+ 
	2 \tilde{E} \tilde{J}  \big( v'(r) w(r) - v(r) w'(r) \big)
	+ u'(r) w(r) - u(r) w'(r) \, .
\end{equation} 
The two equations (\ref{eq:circ1}) and (\ref{eq:circ2}) determine the values of 
$\tilde{E}$ and $\tilde{J}$ for circular orbits in dependence on $r$. To solve these
equations we introduce the functions 
\begin{equation}\label{eq:FGH}
	F(r) = u'(r) v(r) - u(r) v'(r) \, , \quad
	G(r) = w'(r) v(r) - w(r) v'(r) \, , \quad
	H(r) = u'(r) w(r) - u(r) w'(r) \, .
\end{equation} 
Multiplying Eq.(\ref{eq:circ1}) to $H(r) w(r)$ and Eq.(\ref{eq:circ2}) to $u(r)$ and summing them up, we obtain the following quadratic equation 
\begin{equation}\label{eq:quadtJ}
	0 = j^2
	+ 2  \, j \, \dfrac{F(r)}{H(r)}
	- \dfrac{u'(r)}{H(r)} ,
\end{equation} 
for the specific angular momentum $j$ defined as
\begin{equation}\label{eq:j}
	j = \dfrac{\tilde{J}}{\tilde{E}} \, .
\end{equation} 
Solving Eq.(\ref{eq:quadtJ}) with respect to $j$ we get 
\begin{equation}\label{eq:tJ}
	j_{\pm}
	= \dfrac{-F(r) \pm \sqrt{F(r)^2+u'(r) H(r)}}{H(r)} \, .
\end{equation} 
Inserting this into (\ref{eq:circ2}) determines the energy of spinning particle at the circular orbit 
\begin{equation}\label{eq:tE}
	\tilde{E}{}_{\pm}^2 = 
	\dfrac{
		H(r)^2
	}{
		-w'(r) H(r) +2  G(r) \big( -F(r) \pm \sqrt{F(r)^2+u'(r)H(r)} \, \big) \, } \, .
\end{equation} 
If, for a certain radius $r$, the expression under the square-root in 
(\ref{eq:tJ}) and (\ref{eq:tE}) is negative, there are no circular orbits at 
this location. Since
\begin{equation}\label{eq:root1}
	F(r)^2+u'(r)H(r) = \dfrac{\Delta_r^2\,}{r^4 \chi^4 \big( r^3 - \tilde{M} \tilde{S}{}^2 \big)^4}
	\, \psi (r) , 
\end{equation}

\begin{figure}
	\centering
	\begin{subfigure}[h]{.5\textwidth}
		\centering
		\includegraphics[width=.95\linewidth]
		{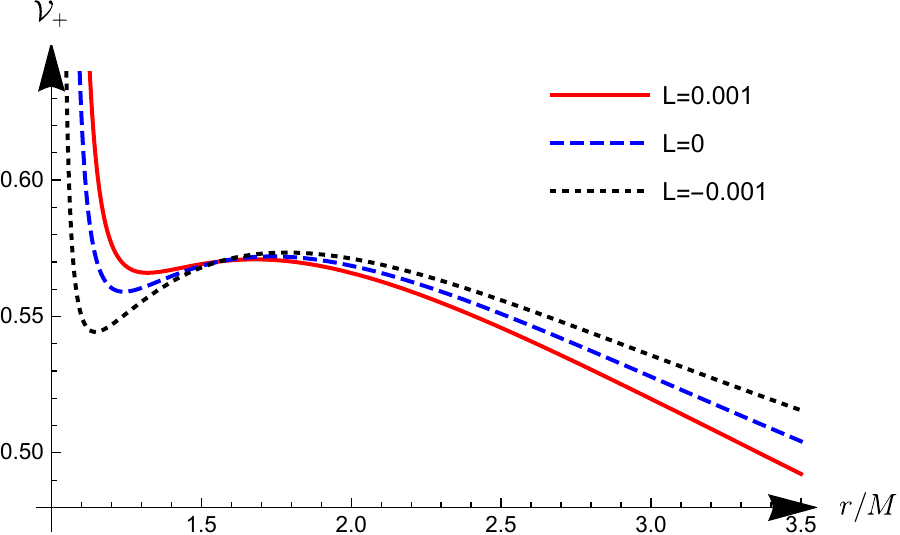}
		\caption{ For zero spin and $a=0.9985$.}\label{fig:1}
	\end{subfigure}%
	\begin{subfigure}[h]{.5\textwidth}
		\centering
		\includegraphics[width=.95\linewidth]
		{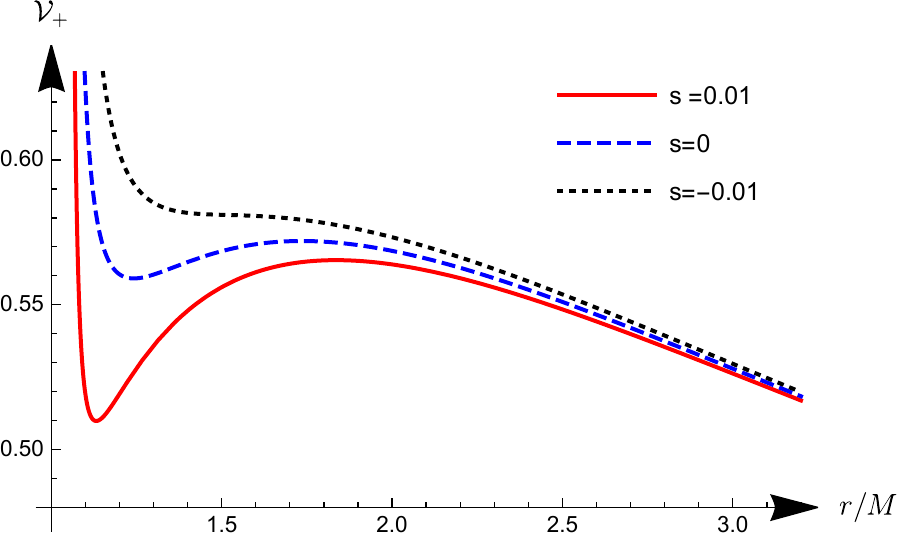}
		\caption{ For zero Cosmological constant and $a=0.9985$.}\label{fig:2}
	\end{subfigure}%
	\caption{The dependence of velocity $\mathcal{V}$ on the radial position of a spinning particle. As one can see, for the negative cosmological constant the minimum value of angular velocity is smaller than zero one. It is opposite for spin: when the spin is positive, the minimum of the angular velocity is lower that for zero and negative spin cases. \label{2} }
\end{figure}

where 
\begin{equation}
	\begin{split}
		M \psi (r)=S^4 \left(9 a^2 M^2+4 \tilde{M}^2 r (\tilde{M}-3 M)\right)+2 a M r^2 S^3 (9 M-2 \tilde{M})+4 a M r^5 S\\+r^4 S^2 \left(-8 \tilde{M}^2+12\tilde{M} M+9 M^2\right)+4 \tilde{M} r^7,
	\end{split} 
\end{equation} 
consequently, one can find the angular velocity in terms of the radius as follows
\begin{equation}\label{eq:Omega3}
	\Omega _{\pm} ^{-1}= 
	\frac{r \left(\pm\sqrt{\psi}+3 M S \left(a S+r^2\right)\right)}{2 a M r S-2 \tilde{M}
		S^2 (\tilde{M}-3 M)+2\tilde{M} r^3}
	+a\,.
\end{equation} 
For a non-spinning case, Eq.\eqref{eq:Omega3} reduces to the well known angular frequency 
\begin{equation}\label{eq:OmegaL}
	\Omega_\pm= \Big(a\pm \sqrt{\dfrac{r^3}{\tilde{M}}}\Big)^{-1}=\Big(a\pm \sqrt{\dfrac{r^3}{M(1-\frac{\Lambda r^3}{M})}}\Big)^{-1}\,.
\end{equation}
Eq.\eqref{eq:Omega3} for zero cosmological constant ($\Lambda=0$) coincides with those obtained in \cite{Khodagholizadeh:2020sex,Hackmann:2014tga}. Therefore one can follow the same argument for determination the co-rotating and counter-rotating motion of test spinning particle, as we did in \cite{Khodagholizadeh:2020sex}. For $a>0$ and far away from the black hole center the $+$ sign corresponds to the co-rotating orbit and $-$ sign to the counter-rotating orbit. 

It is useful to use the locally non-rotating frame (LNRF) of the zero angular momentum observer (ZAMO)\cite{zamo} to define the orbital velocity of a test particle. The LNRF velocity is defined by the orthonormal vierbein for the region, in   which  $\Delta_r > 0$
\begin{equation}
	e_0=
	\dfrac{
		-g_{\varphi \varphi} \partial _t + g_{t \varphi} \partial {\varphi}
	}{
		\sqrt{-g_{\varphi \varphi} \big( g_{t \varphi} ^2 - g_{\varphi \varphi} g_{tt}  \big)}
	} 
	\, , \quad
	e_1 =
	\dfrac{\partial _r}{\sqrt{-g_{rr}}}
	\, , \quad
	e_2 =
	\dfrac{\partial _{\vartheta}}{\sqrt{-g_{\vartheta \vartheta}}}
	\, , \quad
	e_3 =
	\dfrac{\partial _{\varphi}}{\sqrt{-g_{\varphi \varphi}}}
	\, .
	\label{eq:LNRF}
\end{equation}
In the equatorial plane, the orbital velocity $\mathcal{V}$ for a circular motion  with respect to the LNRF is given by
\begin{equation}\label{eq:Vdef}
	u^t \partial _t + u^{\varphi} \partial _{\varphi} =
	N \Big( e_0 + \mathcal{V} \, e_3 \Big) ,
\end{equation}
where $N$ is a scalar factor. For timelike orbits $\mathcal{V}$  is bounded by the speed of light, so its value is restricted 
between $-1$ and 1. Comparing coefficients of $\partial _t$ and 
$\partial _{\varphi}$ in (\ref{eq:Vdef}) allows us to express 
$\Omega = u^{\varphi}/u^t$ in terms of $\mathcal{V}$
\begin{equation}
	\Omega = 
	\dfrac{
		g_{t \varphi} + \mathcal{V} \, 
		\sqrt{  g_{t \varphi} ^2 - g_{\varphi \varphi} g_{tt} }
	}{-g _{\varphi \varphi}}
	\, .
	\label{eq:OmegaV}
\end{equation}
\begin{figure}[h]
	\centering
	\includegraphics[width=.5\linewidth]{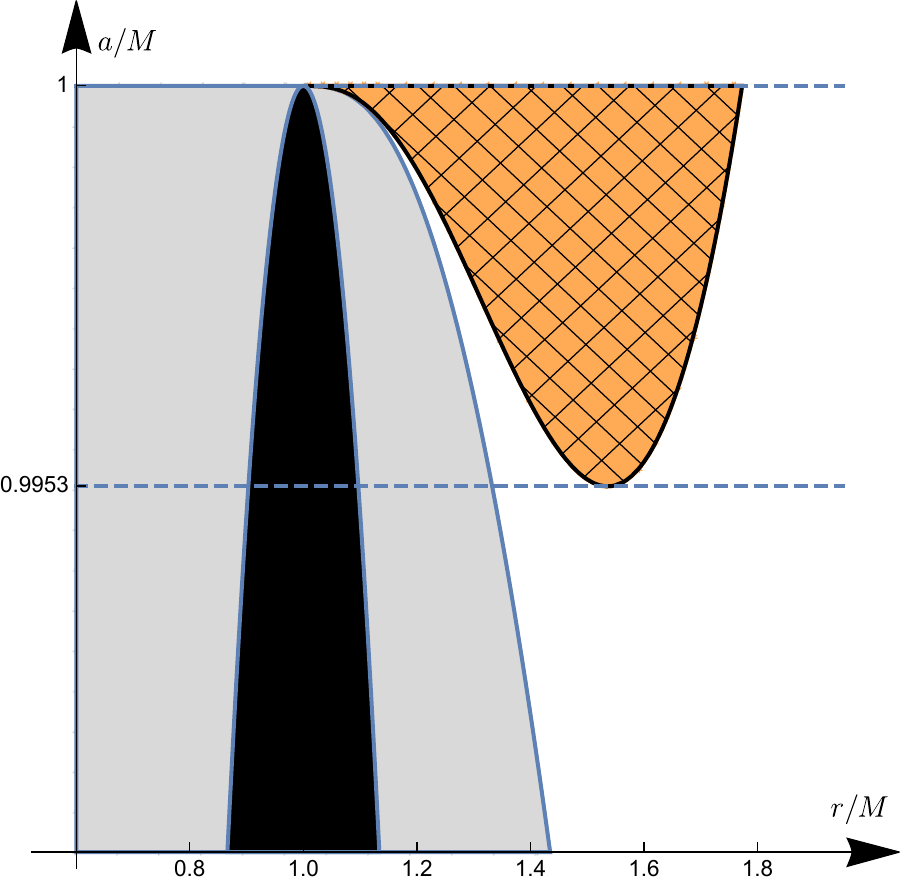}
	\caption{\label{fig:l0s0} The $a-r$ parameter space, where the Aschenbach effect can be observed.
	For $s=0$ and $L=0$ the hatched region shows an allowed values of $a$ and $r$ for the Aschenbach effect,  $\frac{\mathcal{V_+}}{dr} > 0$. The region below the event horizon is colored in black. 
	}
\end{figure}
where
\begin{equation}\label{eq:VOmega}
	\mathcal{V} 
	=
	\dfrac{\big((r^2+a^2)^2-a^2 \Delta_r \big)\Omega-a(2Mr+\frac{\Lambda r^2}{3}(r^2+a^2)}{r^2 \, \sqrt{\Delta_r}}  
	\, .
\end{equation}
In the next step, we investigate the behavior of this orbital velocity in the blac khole background. It should be mentioned that the angular velocity \eqref{eq:VOmega} obtained with the TD condition, at the leading order with respect to the spin  coincides with those of the expansion of angular velocity with the FP condition \eqref{con:FP}. We provide details in the appendix, as well as can be found in  \cite{Khodagholizadeh:2020sex}.

To have an observable effect one needs to be careful about the instability of the circular orbits in which the Aschenbach effect takes place. 
Therefore, it is important to identify the position of the innermost stable circular orbit (ISCO). For that we use a standard effective potential approach. 
Detailed derivation of the effective potential is gien in the Appendix. In order to find the ISCO one needs to solve Eqs.\eqref{vs}. From two equations: $V_{s,\Lambda}(r,a;E,J)=0$ and ${dV_{s,\Lambda}(r,a;E,J)}/{dr}=0$,  we find energy $E$ and angular momentum $J$ in terms of the radial position, black hole spin, cosmological constant and the particle's spin. Using the second derivative of the effective potential, ${d^2V_{s,\Lambda}(r,a;E,J)}/{d^2r}=0$, we find the ISCO radius in terms of $\Lambda, a, s$\footnote{Actually, we also have black hole mass $M$, which we fix to $M=1$ without loss of generality.}. Interestingly we found that the circular orbit, in which angular velocity behaves non-monotonically is placed in the stable region, as we demonstrate by gray regions in the plots which shows an unstable region.

\section{Aschenbach effect in the equatorial plane of Kerr-(A)dS space-time}

\begin{figure}[h]
	\centering
	\begin{subfigure}[h]{0.5\textwidth}
		\centering
		\includegraphics[scale=0.9]{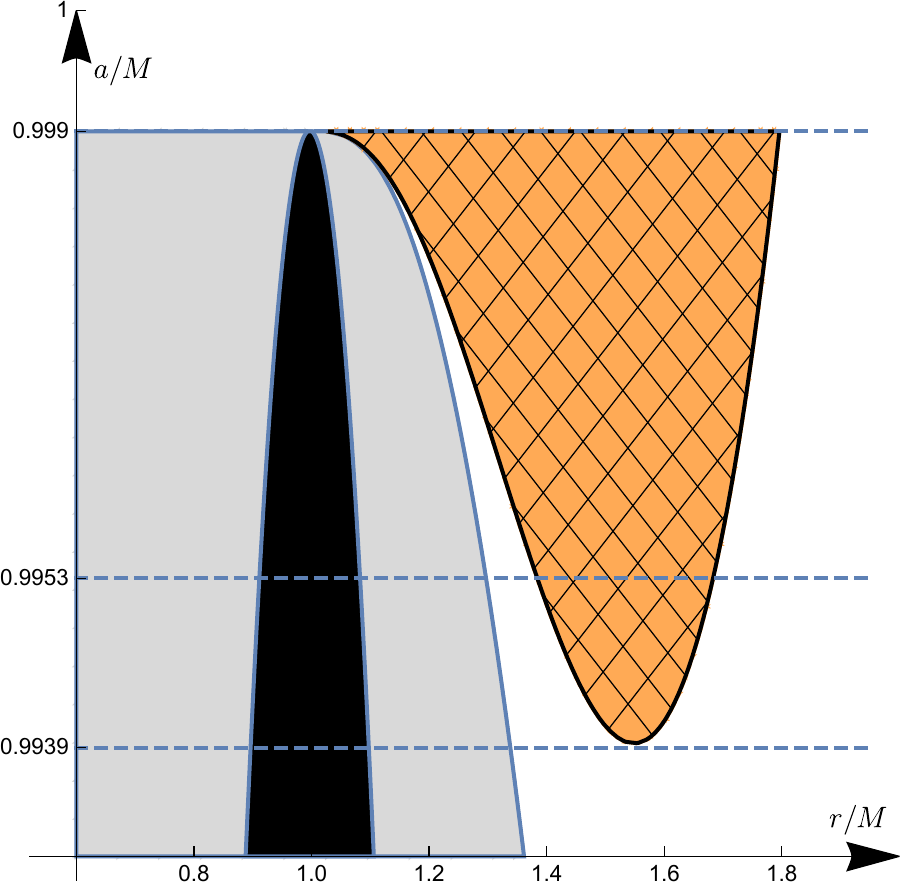}
		\caption{ $s=0$ and $L=-0.001$.}\label{fig:lns0}
	\end{subfigure}%
	~ 
	\begin{subfigure}[h]{0.5\textwidth}
		\centering
		\includegraphics[scale=0.9]{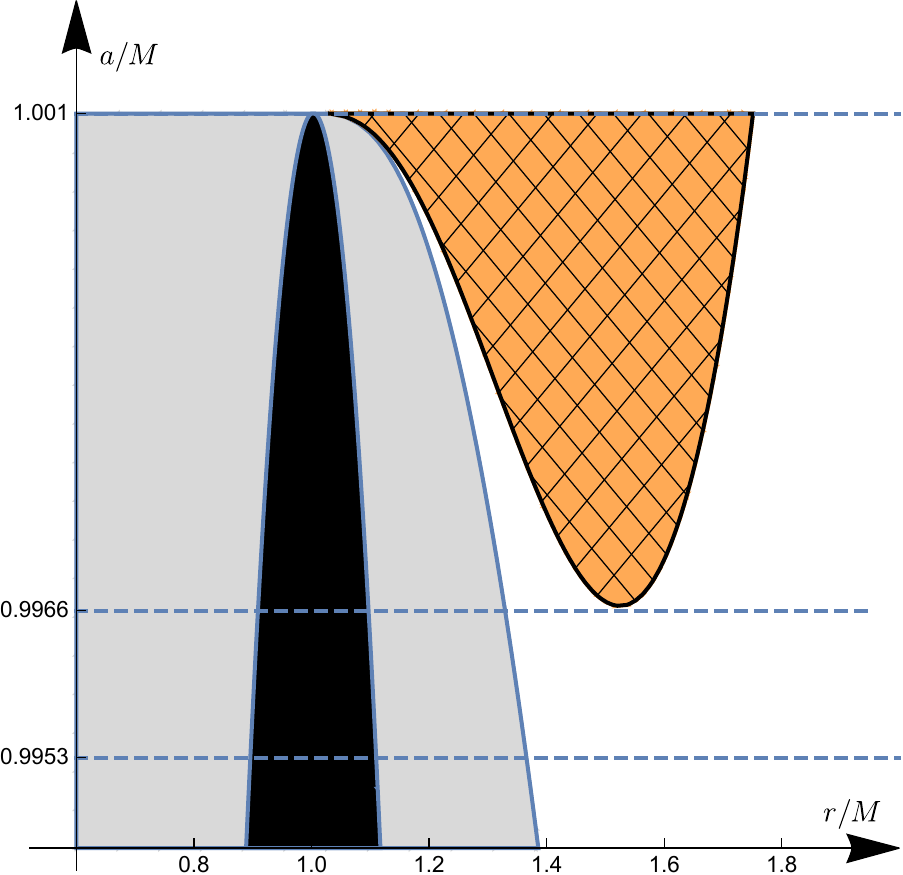}
		\caption{ $s=0$ and $L=0.001$ .}\label{fig:lps0}
	\end{subfigure}
	\caption{\label{fig:zerospin} The $a-r$ parameter space with hatched region, where the Aschenbach effect can be observed for non-spinning particle. The region outside the event horizon corresponds to $\frac{\mathcal{V_+}}{dr}>0$. As one can see for the positive cosmological constant the black hole spin is allowed to be greater than $M$. For the negative value of cosmological constant the critical value of black hole spin is smaller than its value for the zero and positive cosmological constant. The gray region shows unstable parameter's region due to the ISCO consideration.}
\end{figure}

At the limit of zero cosmological constant the spin of a black hole is limited by the black hole mass, i.e, $a \leq M$. However, in the Kerr-(A)dS black hole solution the maximal black hole spin corresponds to ${a}/{M}=1.10092$ for $\Lambda M^2=0.17772$ \citep[see][for details]{Stuchlik:2003dt,Slany:2007gp,Stuchlik:1983baic,Stuchlik:2004prdhle}. 
The orbital velocity of a spinning particle in the Kerr-(A)dS space-time is given by Eq.\eqref{eq:VOmega}. An example of behavior of $\mathcal{V}_+$ in terms of $r$ is shown in Figure \ref{2}. The minimum-maximum structure of effective potential is illustrated in Figure\ref{fig:l0s0} in the $a-r$ plane. As it is clear for zero cosmological constant and zero spin the critical value of the black hole angular momentum is $a_c=0.9953$, see Figure \ref{fig:l0s0}. It is obvious that the maximum-minimum structure for negative cosmological constant is deeper than the other one. The dependence of $a_c$ on the cosmological constant is shown in Figures \ref{fig:lns0} and \ref{fig:lps0}.
The Aschenbach effect at zero spin and zero cosmological constant occurs when the angular momentum of Kerr black hole $a$ is larger than a critical value $a_c=0.9953$   \cite{Aschenbach:2004kj}. The angular velocity of rotating particle in the background with critical angular momentum $a_c$ would satisfy the following conditions\footnote{At special radius. We do not discuss the counter-rotating orbits here, the results are similar to those obtained in \cite{Khodagholizadeh:2020sex}.}
\begin{equation}\label{con:V}
	\dfrac{d \mathcal{V}{}_+}{dr} =0 \, ,  \quad
	\dfrac{d^2 \mathcal{V}{}_+}{dr^2} =0 \,.
\end{equation}
With these conditions one can study the effect of other parameters on the $a_c$ . It has been shown in \cite{Slany:2007gp} that for Kerr-(A)dS space-time the critical value of angular momentum would be a function of cosmological constant. This implies that the non-monotonic behavior of the angular velocity $\mathcal{V}$ depends on $\Lambda$, see Figure \ref{fig:1}. In the next subsection\ref{zeroS}, we find $a_c$ in terms of $\Lambda$ up to the second order using Eq.\eqref{con:V}. 
 Also in \cite{Khodagholizadeh:2020sex} has been illustrated that  for spinning test particle, in the Kerr background, the $a_c$ is control by the spin of test particle, see Figure \eqref{fig:2}.
At the end of this section, we will make a conclusion, at leading order, dependence of $a_c$ with the spin of the test body, $S$, and background cosmological constant, $\Lambda$. 
In order to have an observable effect one should be careful about the instability of the circular orbits in which the Aschenbach effect takes place. Therefore, we have to know the position of the Innermost Stable Circular Orbit (ISCO). To obtain the ISCO we need to solve Eqs.\eqref{vs}, see also \cite{Zhang:2018omr}. Interestingly we found that the circular orbit, in which angular velocity behaves non-monotonically is placed in the stable region, look at the gray regions in the plots, which show an unstable region. 

 It is interesting to point out that Abramowicz and Kluzniak \cite{Abramowicz:2004tm} have shown that the ISCO may occur also in the Newtonian theory of gravity around sufficiently rapidly rotating very dense star. Expanding the Newtonian gravitational potential in terms of spherical harmonics, one can find a region, where the orbits can become unstable, which is however located inside the star if the quadrupole moment is present. Destabilization of orbits can also occur in the presence of higher-order moments \cite{Kluzniak:2001bgr}. Taking arguments from \cite{Abramowicz:2004tm} we raised a question whether the Aschenbach effect can also occur in the Newtonian theory and found that it does not. 


\subsection{Aschenbach effect for non-spinning particle with non-zero $\Lambda$}\label{zeroS}

As mentioned above, for zero-spin particle $\Omega_\pm$ reduces to \eqref{eq:OmegaL}. The effective potential in this case has been studied in details in \cite{Mueller:2007tf,Slany:2007gp}. Here we find the critical value of the black hole spin up to the second order with respect to the cosmological constant.

For small values of $L=\frac{\Lambda\,M^2}{3}$, using \eqref{con:V} we find 
\begin{equation}\label{eq:acL}
	\frac{a_c(L)}{M}=0.99529+0.451253\, L -0.66194 \, L^2 +\,O(L^3).
\end{equation}
Note that in the linear order with respect to $\Lambda$  the results coincide with those obtained in \cite{Slany:2007gp}, see also \cite{ Mueller:2007tf}. In spite of the second order relation in \eqref{eq:acL}, but due to small value of $\Lambda$,   Eq.\eqref{eq:acL}  gives us a similar plot as the one illustrated in \cite{Mueller:2007tf} and also \cite{Slany:2007gp}\footnote{For the meaningful results one needs to be careful about stability of circular orbits, see Appendix B.}.

\subsection{Aschenbach effect for spinning particle with zero $\Lambda$ }\label{zeroL}
As mentioned above, at $\Lambda=0$ all equations, including \eqref{eq:Omega3} and  \eqref{eq:VOmega} coincide with the result in \cite{Khodagholizadeh:2020sex}; For eq. \eqref{eq:VOmega} with $\Lambda=0$ we just have a non-monotonic angular velocity $\mathcal{V}_+$ if $-0.11<\frac{S}{M \mu}<1.05$, see for more details \cite{Khodagholizadeh:2020sex}. 
For small values of $s=\frac{S}{\mu M}$ we have
\begin{equation}\label{eq:acS}
	a_c (s) = \Big( 0.9953 - 0.0517 \, s - 0.0164 \, s^2 + O (s^3) \Big) M  
	\, .
\end{equation}
Note that from Figure 1  in \cite{Khodagholizadeh:2020sex}  the minimum value of $a_c$ is $0.9810$, which corresponds to $s=0.47$.

\subsection{Aschenbach effect: general case}
\begin{figure}[h]
	\centering
	\begin{subfigure}[h]{0.5\textwidth}
		\centering
		\includegraphics[scale=0.9]{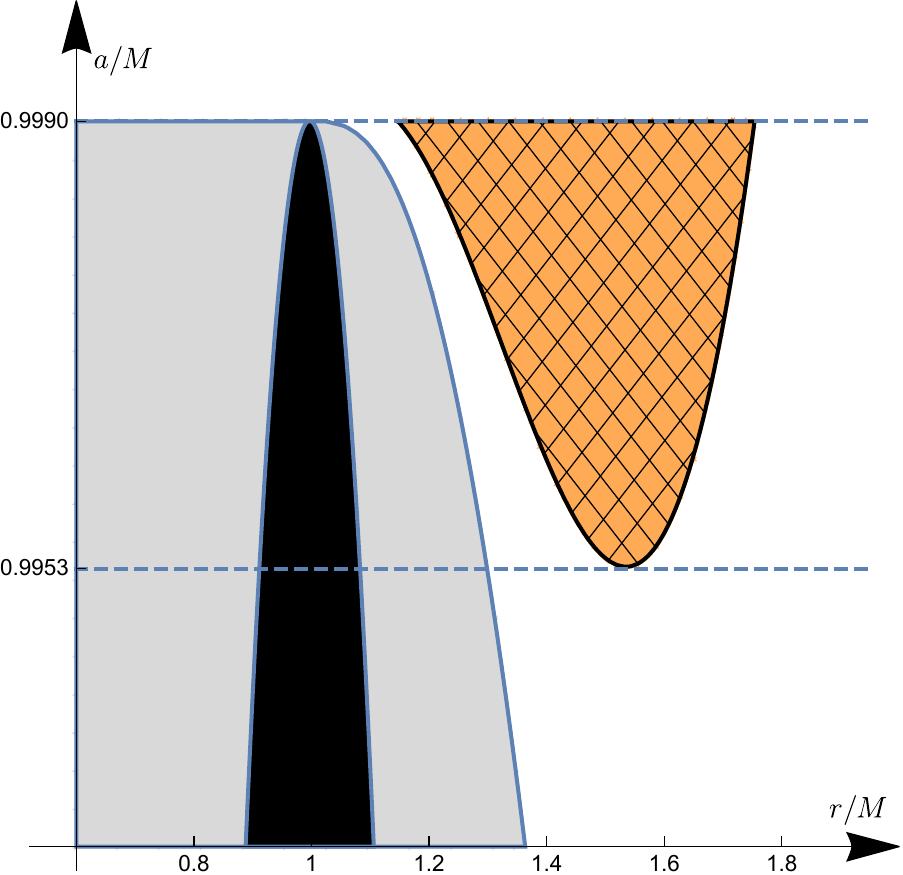}
		\caption{ $s=-0.0038$ and $L=-0.001$ .}\label{fig:l001sn0038}
	\end{subfigure}%
	~ 
	\begin{subfigure}[h]{0.5\textwidth}
		\centering
		\includegraphics[scale=0.9]{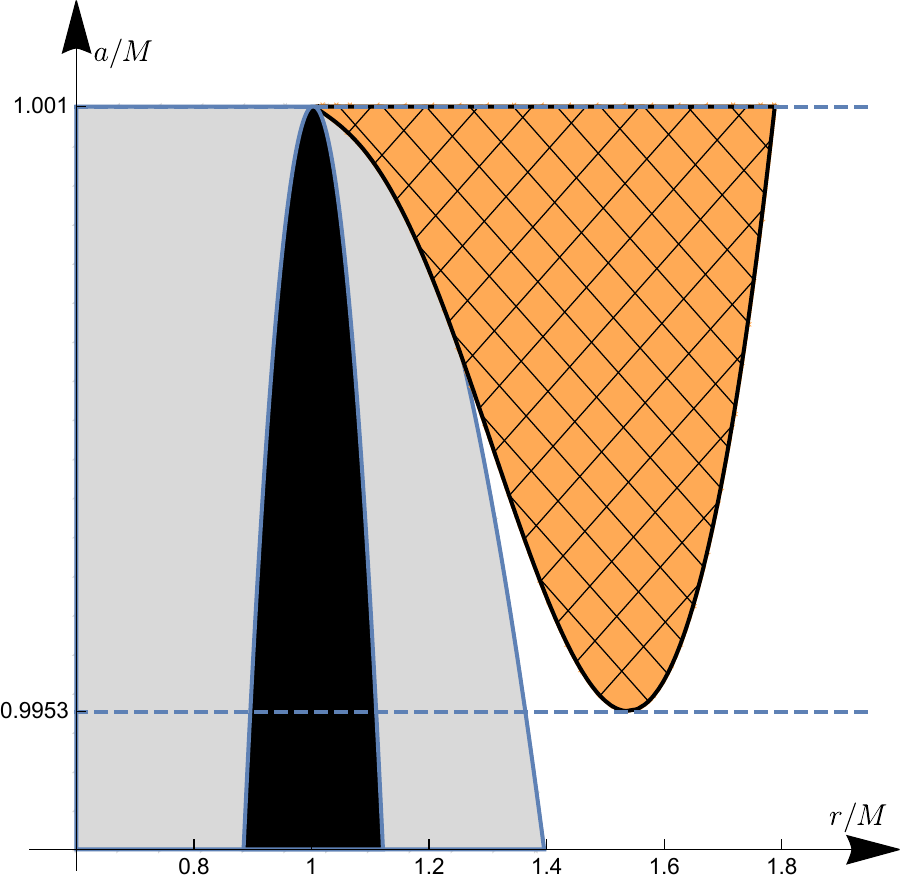}
		\caption{  $s=0.0038$ and $L=0.001$ .}\label{fig:lpsp}
	\end{subfigure}
	\caption{ The $a-r$ parameter space with hatched region, where the Aschenbach effect can be observed in general case. The region outside the event horizon corresponds to $\frac{\mathcal{V_+}}{dr}>0$. The value of the critical spin of the black hole is the same as in the case of non-spinning particle with zero $\Lambda$. 
}\label{fig:sameac}
\end{figure}

For a spinning test particle in the Kerr-(A)dS spacetime the LNRF angular velocity is a function of four parameters, i.e.  $\mathcal{V}=\mathcal{V}(a,S,\Lambda,r)$,  so with a critical conditions \eqref{con:V} one cannot fix $a_c$ as a function of spin $S$ and cosmological parameter $\Lambda$. For a fixed value of black hole's critical spin and using \eqref{con:V}, one can find $\Lambda$ in the terms of a spin $S$ of a test particle or vice versa. 
However, from both \eqref{eq:acS} and \eqref{eq:acL} we can write the critical angular momentum of rotating black hole at leading order with respect to the spin and cosmological constant 
\begin{equation}\label{eq:acSL}
	\frac{a_c(s,L)}{M}=0.99529+0.451253\, L - 0.0517 \, s +... \,\, ,
\end{equation}
where dots stand for higher order terms in $s$ and $L$ and also cross terms $sL$. As mentioned previously, the relation between the spin and cosmological constant could be realized by fixing $a_c$. For example, we can fix ${a_c}/{M}=0.9953$ and look for which spin $S$ and cosmological constant $\Lambda$ we get the same behavior function $\mathcal{V}(r)$, see Figure \ref{fig:sameac}.


We see that for $s=0.0038$, $L=0.001$, 
or $s=-0.0038$ and $L=-0.001$, one can 
obtain the same critical spin for black hole 
as in case of zero $\Lambda$ and non-spinning 
particle, recovering thus, the results in \cite{Khodagholizadeh:2020sex}.

\begin{figure}[h]
	\centering
	\begin{subfigure}[h]{0.5\textwidth}
		\centering
		\includegraphics[scale=0.9]{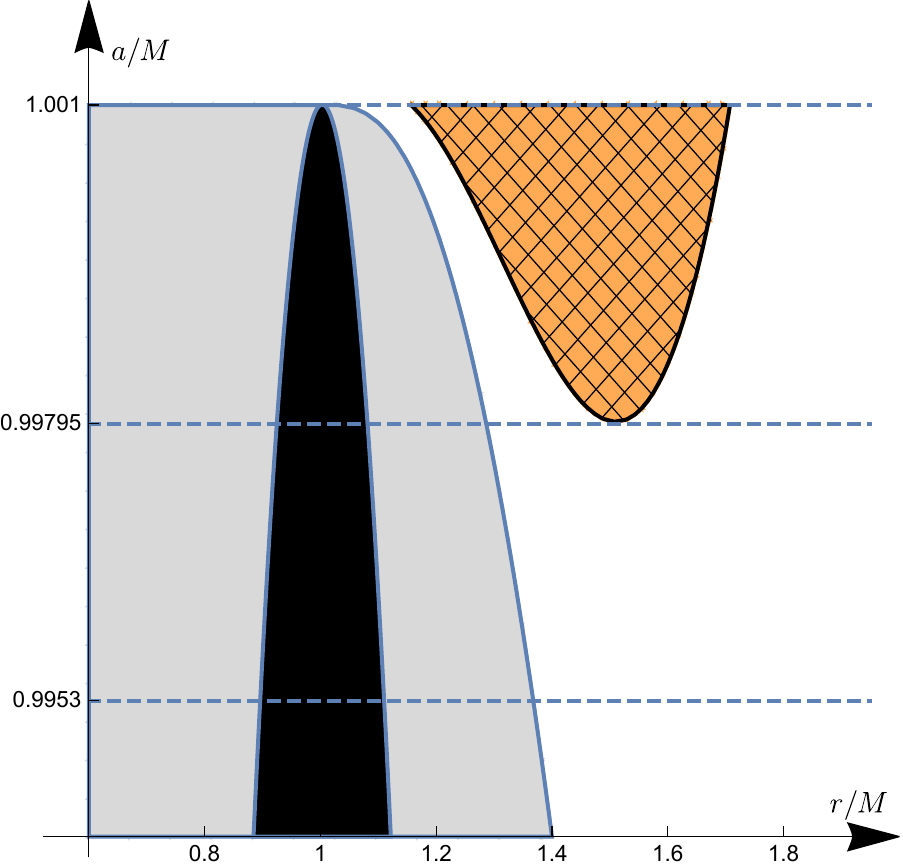}
		\caption{ $s=-0.0038$ and $L=0.001$. }\label{fig:lpsn}
	\end{subfigure}%
	~ 
	\begin{subfigure}[h]{0.5\textwidth}
		\centering
		\includegraphics[scale=0.9]{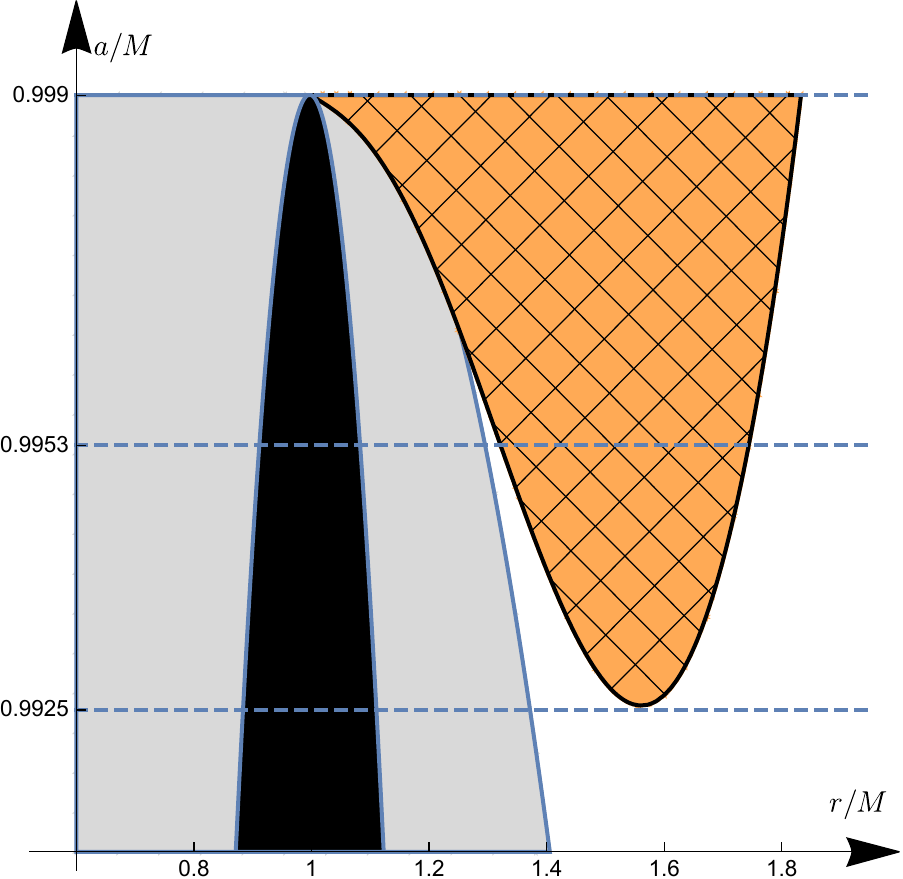}
		\caption{  $s=0.0038$ and $L=-0.001$ .}\label{fig:lnsp}
	\end{subfigure}
	\caption{Example of the $a-r$ parameter space with $a_c$ greater and lower than original pure Kerr value $a_c=0.9953$.
	The hatched regions correspond to the parameter space, where the Aschenbach effect can be observed in general case. 
	}
\end{figure}


\section{Conclusion}
B. Aschenbach in mid 2000's reported a new effect of non-monotonic behavior of the orbital velocity of a test particle in close vicinity of rapidly rotating black hole, if the black hole spin is in the range of $0.9953<a<1$ and orbital radius is below $1.8 M$. Combining the new effect and the observational data related to the quasi-periodic oscillations (QPOs) of the X-ray and near-infrared flux from the Galactic centre supermassive black hole, Aschenbach estimated the spin of the central object to $a=0.99616$ for the mass of $3.3 M_\odot$ to date. Although the current estimates of the mass of the Galactic centre black hole are somewhat different, its spin values are not constrained remaining an open question. Therefore, the Aschenbach effect can provide us a useful tool for the spin estimates of black holes when more data and precise measurements from the immediate vicinity of black holes become available.  The study of the connection of the QPOs phenomena with the Aschenbach effect in the case of a spinning particle moving around Kerr-(A)dS black hole is interesting and important, which can potentially lead to the new predictions related to the effect of the particle's spin and the cosmological constant. We leave such studies for our future works.

In the present paper we have generalized the Aschenbach effect to the case of spinning particle motion around rotating black hole with a cosmological constant. For this case, we derived the MPD equations with the TD boundary condition. Focusing on the motion at the equatorial plane we found the LNRF velocity and the position of the innermost stable circular orbit (ISCO). We have shown that the ISCO radii are located inside the region of the existence of the Aschenback effect also in the case of a spinning particle in the Kerr-(A)dS black hole background. 
We have demostrated that both the particle’s spins $s$ and the cosmological constant $\Lambda$ modify the critical value of the black hole spin $a_c$, for which the Aschenbach effect can be observed; $a_c$ can increase or decrease depending on the signs of $s$ and $\Lambda$. We  found a condition, for which the cosmological constant eliminates the effect of the particle’s spin, which occurs in the cases when $s$ and $\Lambda$ have the same sign. 

One of the important results is that the critical spin $a_c$ is decreasing with increasing an absolute value of the negative cosmological constant $\Lambda<0$ and the positive particle's spin parameter $s$ (see, Figure \ref{fig:lnsp}). Similarly, $a_c$ is increasing towards the extremal black hole when increasing the values of positive $\Lambda>0$ and negative values of the particle's spin $s$ (see, Figure \ref{fig:lpsn}). Interestingly, when both $\Lambda$ and $s$ have the same sign, one can choose their values in such a way that $a_c$ will have the same value as in the pure Kerr case with zero-$\Lambda$ and non-spinning particle.  
For example, for $s=-0.0038$, $L=-0.001$ or $s=0.0038$, $L=0.001$  the black hole's critical spin value $ a_{c}=0.9953$ is obtained (see Figure \ref{fig:l001sn0038}), corresponding to $s=0$ and $L=0$ case. This implies that the particle's spin $s$ can mimic the influence of the cosmological constant in the Aschenbach effect, which may cause a discrepancy in the measurements of $s$ $\Lambda$ and $a_c$.  


\begin{acknowledgments}
	We thank Perlick Volker and Zden\v{e}k Stuchl\'\i{}k for useful comments. 
\end{acknowledgments}

\section*{Appendix-A}

In this appendix we first briefly derive the MPD equations without specific supplementary conditions and then apply TD condition to check the obtained equations \eqref{eq:Omega3}.

\subsection*{Spinning particle in the equatorial plane of the Kerr-(A)dS spacetime}\label{sec:Kerrds}

We specify the background metric to the Kerr metric with a cosmological constant, which reads in standard Boyer-Lindquist coordinates as follows
\begin{equation}\label{metricall0}
	g_{\mu \nu} dx^{\mu} dx^{\nu}
	=\frac{1}{\rho^2 \chi^2}\left[-\Delta_r \left(dt-a \sin^2\theta d\varphi\right)^2+\Delta_\theta \sin^2\theta  \left(a dt-(a^2+r^2) d\varphi\right)^2 \right]+\rho^2\left[\frac{dr^2}{\Delta_r}+\frac{  d\theta^2}{\Delta_\theta}\right], 
\end{equation} 
where
\begin{align}
	\rho^2&=r^2+a^2 \cos^2 \theta, \nonumber\\
	\chi^2&=1+\frac{\Lambda a^2}{3}, \nonumber\\
	\Delta_l&=(r^2+a^2)(1-\frac{\Lambda r^2}{3})-2 M r, \nonumber\\
	\Delta_\theta&=\, (1+\frac{\Lambda a^2 \cos^2\theta }{3}), 
\end{align}
and 
\begin{equation}
	\rho  := r^2+a^2 \mathrm{cos}^2\vartheta
	\, , \quad
	\Delta  = r^2+a^2-2Mr \, .
	\label{eq:rhoDelta}
\end{equation}
Here $M$ is the mass parameter and $a$ is the spin parameter. Both
have the dimension of length. For $a^2 \leq M^2$ we have a black hole whereas for $a^2 > M^2$ we have a naked singularity. 

We consider the Mathisson-Papapetrou-Dixon equations with a 
supplementary condition $ V^{\mu} S_{\mu \nu}=0$, where 
$V^{\mu}$ denotes the 4-velocity field of observer in
circular motion
\begin{equation}
	V^{\mu} \partial _{\mu} = V^t \partial _t + V^{\varphi} \partial _{\varphi}
	\, .
	\label{eq:supplcirc}
\end{equation}
We focus on the circular motion in the equatorial plane, so that 
\begin{equation}
	\vartheta = \pi /2 \, , \quad
	u^{\mu} \partial _{\mu} = u^t \partial _t + u^{\varphi} \partial _{\varphi}
	\, ,
	\label{eq:eqcirc}
\end{equation}
with the spin perpendicular to the equatorial plane,
\begin{equation}
	S^{\mu} \partial _{\mu} = S^{\vartheta} \partial _{\vartheta} 
	\, , \quad
	S^{\vartheta} = - \, \dfrac{S}{r} \, .
	\label{eq:spinperp}
\end{equation}
Here $S$ is a constant of motion, satisfying the condition 
\begin{equation}
	g_{\mu \nu} S^{\mu} S^{\nu} = - S^2 \, ,
	\label{eq:S0}
\end{equation}
that may be positive or negative. We have $aS>0$ if the spin of the particle is parallel to the spin of the black hole and $aS<0$ if it is antiparallel.  

Under these assumptions, evaluating all components of the MPD equation (\ref{eq:MPD3}) yields
\[
p^r =0 \, , \quad p^{\vartheta} = 0 \, , 
\]
\begin{equation}
	- S u ^{\varphi} V ^{\varphi}   + \dfrac{\tilde{M}(r) S}{r^3} \big( u^t-a u^{\varphi} \big) 
	\big( V^t - a V^{\varphi} \big) = 
	\big( p^t - a p^{\varphi} \big) u^{\varphi}  - p ^{\varphi} \big(u^t-a u^{\varphi} \big)
	\, .
	\label{eq:Eq1}
\end{equation}
Similarly, from  (\ref{eq:MPD2}) we find
\[
\dfrac{dp^t}{ds}= 0  \, , \quad
\dfrac{dp^{\varphi}}{ds}= 0  \, , 
\]
\[
\tilde{M} \, r^2 \big( p^t - a p^{\varphi} \big)  \big( u^t - a u^{\varphi} \big) 
-r^5  p^{\varphi}  u^{\varphi} 
\]
\begin{equation}
	=    -3 MS a \big( u^t- a u^{\varphi} \big) \big( V^t-a V^{\varphi} \big) + 
	S r^2 \Big( (3M-\tilde{M}) ( u^t- a u^{\varphi} ) V^{\varphi} +\tilde{M} u^{\varphi} (V^t-a V^{\varphi} ) \Big) 
	\, .
	\label{eq:Eq2}
\end{equation}

The FD condition follows to the relation: $V^{\mu}=u^\mu$, while one can apply the TD condition if $V^{\mu} = p^{\mu} / \mu$. 

\subsection*{Tulczyjew-Dixon condition}
If the Tulczyjew-Dixon supplementary condition $V^{\rho} = p^{\rho} / \mu$ 
is imposed, $\mu$ is a constant of the motion and it is convenient to characterize the
particle's spin by the dimensionless parameter
\begin{equation}
	s = \dfrac{S}{ \, \mu} \, .
	\label{eq:defs}
\end{equation}
Eqs. (\ref{eq:Eq1}) and (\ref{eq:Eq2}) satisfy to 
\begin{equation}
	\dfrac{p^t - a p^{\varphi}}{p^{\varphi}} =
	\dfrac{
		u^t- a u^{\varphi}  -  S u^{\varphi} 
	}{
		u^{\varphi} - \dfrac{\tilde{M} S}{r^3} \big( u^t - a u {\varphi}\big)
	}
	\, ,
	\label{eq:Eq1TD}
\end{equation} 
\begin{equation}
	\dfrac{p^t - a p^{\varphi}}{p^{\varphi}}   = 
	\dfrac{
		r^5 u^{\varphi} + S  r^2(3M -\tilde{M}) ( u^t - a u^{\varphi} ) 
	}{
		\tilde{M}	r^2   ( u^t-a u^{\varphi} )   + 3 a S M ( u ^t - a u^{\varphi} )  
		-S\tilde{M}  r^2  u ^{\varphi} 
	}
	\, , 
	\label{eq:Eq2TD}
\end{equation} 
\begin{equation} 
	\Omega = \dfrac{u^{\varphi}}{u^t} .
	\label{eq:Omega}
\end{equation}
Equating the right-hand sides of (\ref{eq:Eq1TD}) and (\ref{eq:Eq2TD}) yields 
\begin{equation} 
	\left( \tilde{M}  + \dfrac{3a S M}{r^2 } + \dfrac{\tilde{M} S^2}{r^3}(3M-\tilde{M}) \right) 
	\big(\Omega ^{-1} - a \big )^2
	- 3 S M  \left( 1 + \dfrac{a S}{r^2}\right) \big(\Omega ^{-1} - a \big )
	- r^3 + S^2 \tilde{M} = 0 \, .
	\label{eq:quadTD}
\end{equation}
This is a quadratic equation in $(\Omega ^{-1} -a )$ with solutions 
\begin{equation} 
	\Omega _{\pm} ^{-1} - a 
	= 
	\frac{r \left(\pm\sqrt{D}+3 M S \left(a S+r^2\right)\right)}{2 a M r S-2 \tilde{M}
		S^2 (\tilde{M}-3 M)+2\tilde{M} r^3}\, ,
	\label{eq:OmegaTD}
\end{equation}
where
\begin{equation}
	D = S^4 \left(9 a^2 M^2+4 \tilde{M}^2 r (\tilde{M}-3 M)\right)+2 a M r^2 S^3 (9 M-2 \tilde{M})+4 a M r^5 S+r^4 S^2 \left(-8 \tilde{M}^2+12\tilde{M} M+9 M^2\right)+4 \tilde{M} r^7
	\, .
	\label{eq:defD}
\end{equation}
Note that because of the normalization condition (\ref{eq:proper}),
we have
\begin{equation}
	g_{tt} \Omega ^{-2} + 2 g_{t \varphi} \Omega ^{-1}
	+ g_{\varphi \varphi} = \dfrac{1}{\big( u^{\varphi} \big)^2} 
	\, .
	\label{eq:uphi} 
\end{equation}
After inserting the metric coefficients the condition of 
$1/ \big( u ^{\varphi} \big) ^2 > 0$ requires that 
\begin{equation}
	\frac{1}{\chi}(\Big( (1-L(r^2+a^2))- \dfrac{2M}{r} \Big) \big( \Omega ^{-1} - a \big) ^2 
	+ 2 a \big( \Omega ^{-1} - a \big) 
	-  r^2 )  > 0 \, .
	\label{eq:sublum}
\end{equation}
This inequality ensures that the 4-velocity of the particle is 
timelike, i.e., the motion is subluminal. If at a certain radius $r$, the discriminant $D$ defined in 
(\ref{eq:defD}) is negative, then there is no real solution at this radial value. If $D$ is non-negative, there may be 
two solutions (typically one co-rotating and the other one is counter-rotating).  
Whether there is a solution or not, depends on whether (\ref{eq:sublum})  
is satisfied to one of $\Omega = \Omega _+$ or $\Omega = 
\Omega _-$, or not.

Using the definition of the orbital velocity in the locally non-rotating frame (LNRF), one can rewrite the equation (\ref{eq:VOmega}) as follows 
\begin{equation}\label{eq:VOmega0}
	\mathcal{V} 
	=
	\dfrac{\big((r^2+a^2)^2-a^2 \Delta_l \big)\Omega-a(2Mr+L r^2(r^2+a^2)}{r^2 \, \sqrt{\Delta_l}}  
	\, .
\end{equation}
Substituting $\Omega=\Omega _{\pm}$ from (\ref {eq:OmegaTD}), we obtain  
two solutions for the orbital velocity, $\mathcal{V}_\pm$.

For the existence of an orbit with velocity $\mathcal{V}_+$ (or $\mathcal{V}_-$,
respectively) at the radius $r$, it is necessary and sufficient that $D$ is 
non-negative and that $\big| \mathcal{V}{}_+ \big| < 1$ (or 
$\big| \mathcal{V}{}_- \big| < 1$, respectively).

Far away from the center, (\ref{eq:OmegaTD}) and (\ref{eq:VOmega0}) can 
be approximated as
\begin{equation}
	\Omega{}_{\pm}
	= 
	\, \pm \, \dfrac{\sqrt{\tilde{M}}}{\sqrt{r^3}} 
	\Big( 1 + O \big( (M/r)^{1/2} \Big)
	\, ,
	\label{eq:asyOmega}
\end{equation}
\begin{equation}
	\mathcal{V}{}_{\pm}
	= 
	\pm \dfrac{\sqrt{\tilde{M}}}{\sqrt{r}} \Big( 1 + O \big( (M/r)^{1/2} \Big)
	\, .
	\label{eq:asyV}
\end{equation} 
From these equations, we see that for any choice of $a$ and $S$,
there are two circular orbits at all sufficiently large radius values;
in Eq.(\ref{eq:asyV}), the sign "$+$" refers to an orbit with positive $\mathcal{V}$, 
i.e., a particle moving in a positive $\varphi$ direction with respect to 
the ZAMO, whereas the the sign "$-$" refers to an orbit with negative 
$\mathcal{V}$, i.e., a particle moving in a negative $\varphi$ direction
with respect to the ZAMO\footnote{Note that $\tilde{M}>0$, which is always true for the negative $\Lambda$. 
}. This means that for $a>0$ the "$+$" orbit
is co-rotating and the "$-$" orbit is counter-rotating; for $a<0$ it is vice versa. Far away from the center, $\mathcal{V}{}_{+}$ goes
monotonically to zero from above and $\mathcal{V}{}_{-}$ goes
monotonically to zero from below. It is the subject of this paper to 
investigate if and how this monotonic behavior changes closer to 
the central object.  As one can see $\mathcal{V}{}_+$ and 
$\mathcal{V}{}_-$ may change signs, so it is not always 
true that for $a>0$ the "$+$" orbit is co-rotating and the "$-$" orbit
is counter-rotating. Also, $\big| \mathcal{V}{}_+ \big|$ and 
$\big| \mathcal{V}{}_- \big|$ may become greater than 1;  in the 
regions where this happens the corresponding orbit does not
exist at all.

\section*{Appendix-B}
In this appendix we derive effective potential for a spinning test body in a Kerr-(A)dS background. The approach is the same as presented in \cite{Saijo1998}. 
{\bf{ 
Our results in the Kerr black hole limit are in accordance with \cite{Abram79}, which is historically the first paper on spinning particles orbiting Kerr black hole. 
 For the non-spinning test particles, please see \cite{Kerner:2001cw,Abramowicz:2004tm} and references therein.}} 

For the metric (\ref{metricall}), one can define the  orthonormal basis as follows
\begin{align} \label{kerrads-veilbein}
	&  \bm{e}^{(0)} = \frac{\sqrt{\Delta_r}}{\sqrt{\Sigma \chi^2}}\Big(dt-a\sin^2\theta d\phi\Big) ~,
	&& \bm{e}^{(1)} = \frac{\sqrt{\Sigma}}{\sqrt{\Delta_r}}\,dr ~, \nonumber\\
	&  \bm{e}^{(3)} = \frac{\sin\theta \sqrt{\Delta_\theta}}{\sqrt{\Sigma \chi^2}}\Big(-a\,dt+(r^2+a^2)d\phi\Big) ~, 
	&& \bm{e}^{(2)} = \frac{\sqrt{\Sigma}}{\sqrt{\Delta_\theta}}\,d\theta ~.
\end{align}
The two conserved quantities in this basis correspond to
\begin{subequations}
	a time-like Killing vector
	\begin{equation}
		\bm{\xi}=-\left(\frac{\sqrt{\Delta_r}}{\sqrt{\Sigma \chi^2}} \bm{e}^{(0)}+\frac{a\sin\theta \sqrt{\Delta_\theta}}{\sqrt{\Sigma\chi^2}} \bm{e}^{(3)}\right), 
	\end{equation}
	and an axial Killing vector
	\begin{equation}
		\bm{\kappa}=a \frac{\sqrt{\Delta_r}}{\sqrt{\Sigma\chi^2}} \sin^2\theta\, \bm{e}^{(0)}+\frac{(a^2+r^2)\sin\theta\sqrt{\Delta_\theta} }{\sqrt{\Sigma \chi^2}}\, \bm{e}^{(3)} \, . 
	\end{equation}
\end{subequations}
Now we fix the plane of the motion to the equatorial one, i.e., $\theta=\frac{\pi}{2}$. The non-vanishing spin component of a spin vector is 
\[s^{(a)}=-\frac{1}{2\mu^2}\epsilon^{(a)}_{(b)(c)(d)} p^{(b)} S^{(c)(d)} ,\] 
%
\begin{equation}
	s^{(2)}=-s, 
\end{equation}
which is a constant\footnote{In this appendix we use $S^{\mu\nu}S_{\mu\nu}=-2\mu^2 s^2$ and following \cite{Saijo1998}, we normalize orbital  affine parameter $\tau$ such that $p^\mu u_\mu=1$ }. This means that
\begin{align}
	S^{(1)(0)}&=- s p^{(3)}\,,\notag\\
	S^{(0)(3)}&= s p^{(1)}\, ,\notag\\
	S^{(1)(3)}&= s p^{(0)}\,,
\end{align}
So the energy and the angular momentum can be written as follows
\begin{subequations}
	\begin{equation}
		\chi\,	E=\frac{\sqrt{\Delta_r}}{r}p^{(0)}+\frac{a r+ \tilde{M} s}{r^2}p^{(3)},
	\end{equation} 
	\begin{equation}
		\chi\,\, J=\frac{\sqrt{\Delta_r}}{r}(a+s)p^{(0)}+[\frac{a s(\tilde{M}+r)}{r^2}+\frac{r^2+a^2}{r}]p^{(3)},
	\end{equation}
\end{subequations}
where we define $\tilde{M}= M(1-\frac{\Lambda r^3}{3\, M})$, as mentioned before. 
The relation between $u^{a}$ and $p^{a}$ can be solved using \eqref{eq:mom} 
\begin{eqnarray}\label{v-p-kerrds}
	\mu	u^{(0)}&=&N^{-1}(1-\frac{\tilde{M} s^{2}}{r^{3}})p^{(0)},\nn\\
	\mu	u^{(1)}&=&N^{-1}(1-\frac{\tilde{M}s^{2}}{r^{3}})p^{(1)},\nn\\
	\mu	u^{(3)}&=&N^{-1}(1+\frac{(3M-\tilde{M})s^{2}}{r^{3}})p^{(3)},
\end{eqnarray}
where
\begin{equation}
	N=(1-\frac{Ms^{2}}{r^{3}}[\frac{\tilde{M}}{M}+3\frac{(p^{(3)})^2}{\mu^2}]).
\end{equation}
Eqs. \eqref{v-p-kerrds} are different from  what is shown in \cite{Zhang:2018ocv}. 
With definition of $u^\mu=\dfrac{dx^\mu}{d\tau}$ and $u^{(a)}=u^\mu e^{(a)}_\mu$ we have
\begin{eqnarray}
	u^{(0)}&=&\sqrt{\frac{\Delta_{r} }{\Sigma \chi^2 }}(\frac{dt}{d \tau }-asin^2\theta  \frac{d\varphi }{d \tau }), \nn\\
	u^{(1)}&=&\sqrt{\frac{\Sigma }{\Delta _r}} \frac{dr}{d \tau },\nn\\
	u^{(3)}&=&\sqrt{\frac{\Delta _\theta}{\Sigma \chi^2}}sin\theta(-a\frac{dt}{d \tau }+(r^2+a^2)\frac{d\varphi }{d \tau }).
\end{eqnarray}
so that for $\theta=\pi/2$ we have
\begin{eqnarray}
	\frac{dt}{\chi\,d\tau}&=&\frac{p^{(0)} \left(a^2+r^2\right) \left(r^3-M s^2\right)+a p^{(3)} \sqrt{\Delta _r} \left(r^3-s^2 (\tilde{M}-3 M)\right)}{N r^4 \sqrt{\Delta _r}}\\
	\frac{dr}{d\tau}&=&\frac{\sqrt{\Delta _r}(1-\frac{\tilde{M} s^2}{r^3})p^{(1)}}{N r}, \label{3.30}\\
	\frac{d \varphi}{\chi d\tau}&=&\frac{a p^{(0)} \left(r^3-M s^2\right)+p^{(3)} \sqrt{\Delta _r} \left(r^3-s^2 (\tilde{M}-3 M)\right)}{N r^4 \sqrt{\Delta _r}}, 
\end{eqnarray}
\begin{equation}
	\begin{split}
		& \Sigma_{s,\Lambda} \Lambda_{s,\Lambda}\frac{ \dot t}{\chi^2} =a\left( 1 +\frac{3 M s^2}{r \Sigma_{s,\Lambda}}\right)\left[ J -(a+s)E \right] +\frac{r^2 +a^2}{\Delta_r}P_{s,\Lambda},\\
		& \Sigma_{s,\Lambda} \Lambda_{s,\Lambda} \frac{\dot \varphi}{\chi^2} =\left( 1 +\frac{3 M s^2}{r \Sigma_{s,\Lambda}}\right)\left[ J -(a+s)E \right] +\frac{a}{\Delta_r}P_{s,\Lambda}.\\
		& \Sigma_{s,\Lambda} \Lambda_{s,\Lambda} \dot r=-\sqrt{R_{s,\Lambda}} ,
	\end{split}
	\label{spin-lambda-eqs}
\end{equation}
where
\begin{equation}
	\begin{split}
		& \Sigma_{s,\Lambda}= r^2 \left(1 -\frac{\tilde{M}s^2}{r^3}\right),\\
		& \Lambda_{s,\Lambda}= 1 - \frac{s^2}{r^3}(\tilde{M}+\frac{3 M  r^2 [-(a + s) E +J]^2}{\Sigma_s^2}),\\
		&R_{s,\Lambda} = P_s^2 - \Delta_r \left\{ \frac{\Sigma_s^2}{r^2} + [-(a + s) E +J]^2 \right\}, \\
		&P_{s,\Lambda} = \left[r^2 + a^2 + \frac{a s (r + \tilde{M})}{r} \right]  E  -   \left(a + \frac{\tilde{M} s}{r} \right) J.
	\end{split}
\end{equation}
For zero spin Eqs. \eqref{spin-lambda-eqs}, satisfy the spin-less geodesic equation  \cite{Kraniotis:2005zm,Frolov:2017kze} and for zero cosmological constant these equations reduce to the results obtained in \cite{Saijo1998}. 

\begin{figure}[h]
	\centering
	\includegraphics[width=.5\textwidth]
	{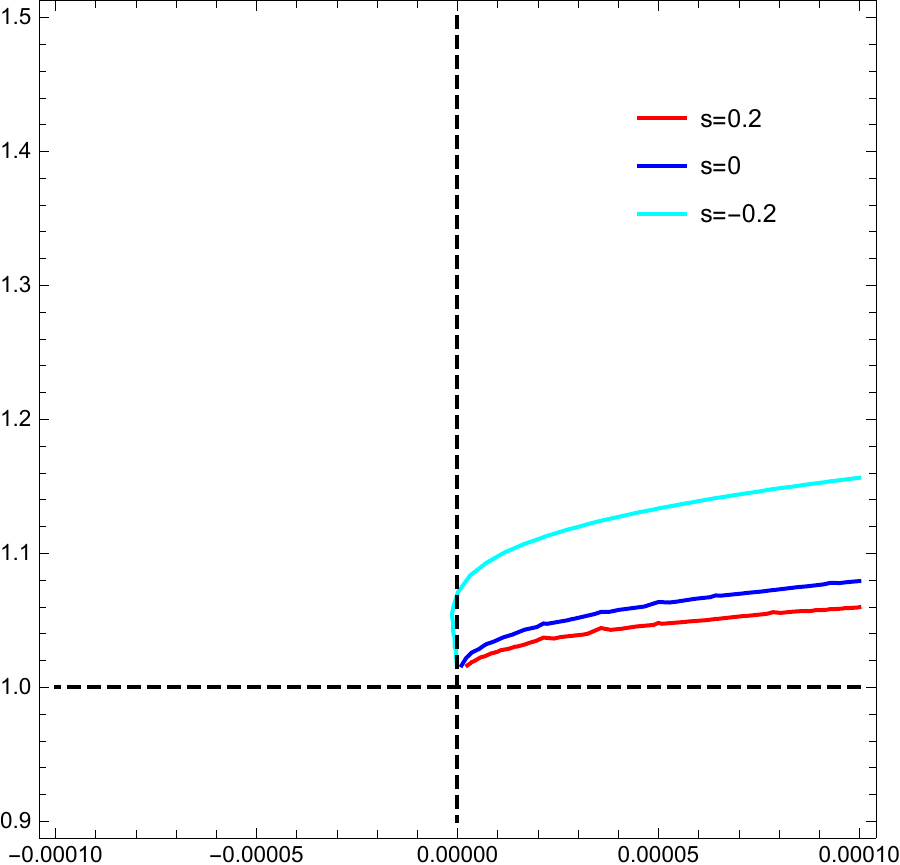}
	\caption{The innermost stable circular orbit for co-rotating test body in term of $L$, around a black hole with $a/M=0.9999$. In this case, $r_{ISCO}$ exists only for  $\Lambda\geq 0 $, except for negative spin values. }\label{fig:ISCO4a.9999}
\end{figure}

One can rewrite the equation for radial motion as
\begin{equation}
	(\Sigma_{s,\Lambda} \Lambda_{s,\Lambda} \dot r)^2=\alpha_{s,\Lambda} E^2 -2\beta_{s,\Lambda} E +\gamma_{s,\Lambda},
\end{equation}
which is the same as in  \cite{Saijo1998,Jefremov:2015gza}, corresponding to the zero $\Lambda$ case. One can define
the effective potential as \footnote{The form of this effective potential is slightly different from that presented in \cite{Zhang:2018omr}, however the results from Eqs.\eqref{vs} for the ISCO radii are the same.}
\begin{equation}\label{eff}
	V_{s,\Lambda}(r;J,E)= \frac{1}{\Sigma_{s,\Lambda}^2 \Lambda_{s,\Lambda}^2} (\alpha_{s,\Lambda} E^2 -2\beta_{s,\Lambda} E +\gamma_{s,\Lambda}) , 
\end{equation}
where 
\begin{equation}
	\begin{split}
		&\alpha_{s,\Lambda} = \left[r^2 + a^2 + \frac{a s (r + \tilde{M})}{r} \right]^2 - \Delta_r (a + s)^2,\\
		&\beta_{s,\Lambda} = \left[ \left(a + \frac{\tilde{M} s}{r} \right) \left(r^2 + a^2 + \frac{a s (r + \tilde{M})}{r} \right) - \Delta_r (a + s) \right]J\\
		&\gamma_{s,\Lambda} = \left(a + \frac{\tilde{M} s}{r} \right)^2 J^2 - \Delta_r \left[r^2 \left(1 - \frac{\tilde{M} s^2}{r^3}\right)^2 + J^2 \right].
	\end{split} \label{abc-spinlambda}
\end{equation}
The innermost stable circular orbit (ISCO) can be found by solving the following equations
\begin{equation}\label{vs}
	\left\{
	\begin{aligned}
		V_{s,\Lambda} &= 0 \, ,\\
		\frac{dV_{s,\Lambda}}{dr} &=0 \, ,\\
		\frac{d^2 V_{s,\Lambda}}{dr^2} &=0.
	\end{aligned}
	\right.
\end{equation}
From $V_{s,\Lambda}=0$ and ${dV_{s,\Lambda}}/{dr} =0$ we can find the energy $E$ and angular momentum $J$ of a test body in terms of radius, black hole spin, particle's spin and cosmological constant\footnote{The first paper that discussed the question of the ISCO location, and changes in the	orbits stability due to the spin of a test body in Kerr background is \cite{Abram79}. Their perturbation results are in agreement with our numeric analysis at $\Lambda=0$.}. The case with $J>0$ stands for co-rotating orbits and $J<0$ for counter-rotating orbits, see \cite{Abram79,Jefremov:2015gza}. By inserting them back in the effective potential, the equation $\frac{d^2 V_{s,\Lambda}}{dr^2}=0$ gives us the ISCO radius. The ISCO radius depends on both cosmological constant and spin of a test body. In case of $\Lambda=0$ for co-rotating orbit, the ISCO is bounded to $ M<r<6 M$ and for counter-rotating one $ 6M<r<9 M$ if $0<a<M$, \citep[see][for more details]{Abram79,Jefremov:2015gza,Howes}. For near-extremal Kerr black hole, i.e., $a\rightarrow 0$ and co-rotating particle we have $r_{ISCO}\rightarrow M$, as can be seen in Fig.\eqref{fig:ISCO4a.9999}. The figure shows that $r_{ISCO}$ is not affected by  negative $\Lambda$\footnote{This is not true for counter-rotating particle.}.

\begin{figure}[h]
	\centering
	\includegraphics[width=.5\textwidth]
	{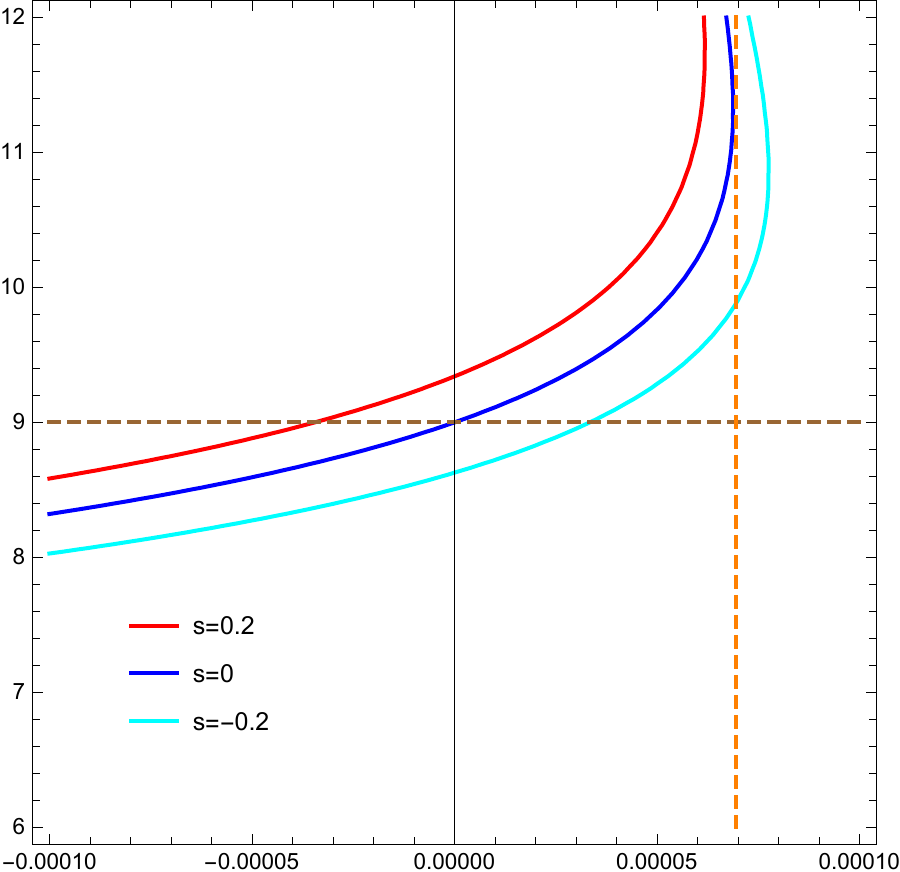}
	\caption{ISCO radius in term of $L$ for counter-rotating test body with $a/M=1$. The counter-rotating case corresponds to $r_{ISCO}/M=9$ at $\Lambda=0$. For $ L > 0.0000695$, orange dashed line, $r_{ISCO}$ does not exist in case of zero spin for $a/M=1$.}\label{fig:ISCO-counter4a=1}
\end{figure}

In Fig. \eqref{fig:ISCO-counter4a=1}, we demonstrate the behavior of the ISCO radius in term of cosmological constant.For counter-rotating particle  in the background with $\Lambda=0$, ISCO radius for  counter-rotating test body is located at $r_{ISCO}=9M$, but for positive spin $r_{ISCO}>9M$ and for negative spin $r_{ISCO}<9M$. These result is in agreement with \cite{Abram79} and also \cite{Jefremov:2015gza}. 

For slowly rotating black hole i.e., $a\ll M$, in case of co-rotating particle we have $r_{ISCO}<6M $ and for counter-rotating one we have $r_{ISCO}<6M $ for zero cosmological constant, see the following figures and also \cite{Abram79,Jefremov:2015gza}.
\begin{figure}[h]
	\centering\includegraphics[width=.5\textwidth]
	{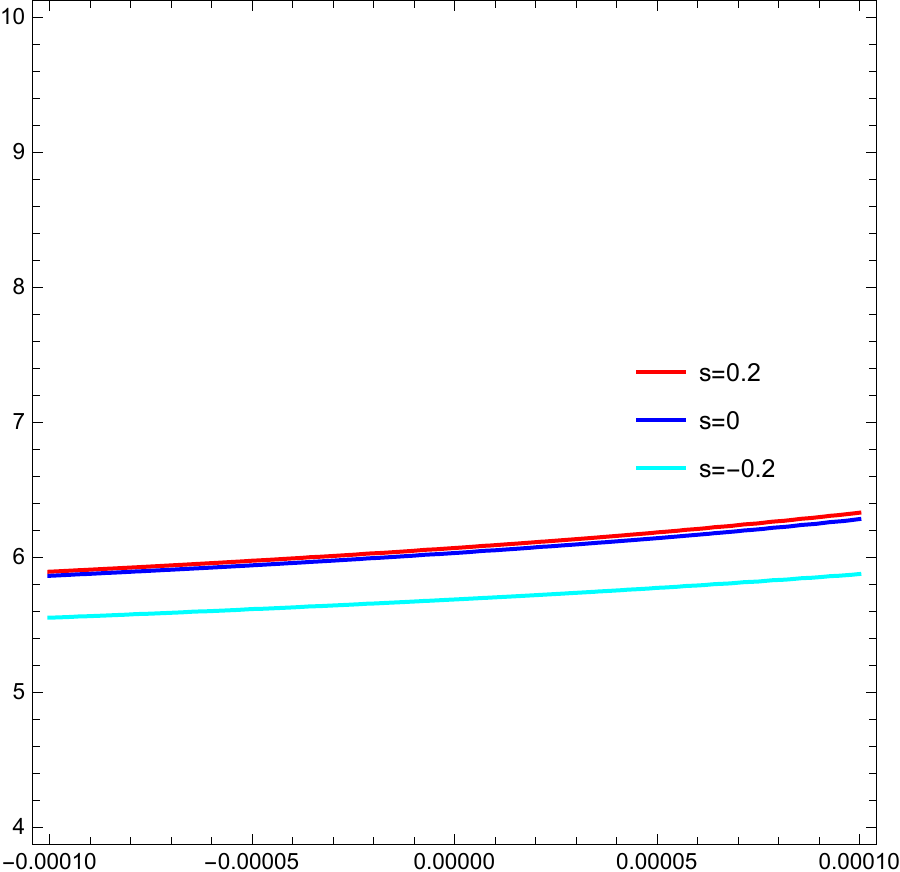}
	\caption{ISCO radius in term of $L$, for counter-rotating test body with $a/M=0.01$.}\label{fig:ISCO-counter4a=.1}
\end{figure}

\begin{figure}[h]
	\centering
	\includegraphics[width=.5\textwidth]
	{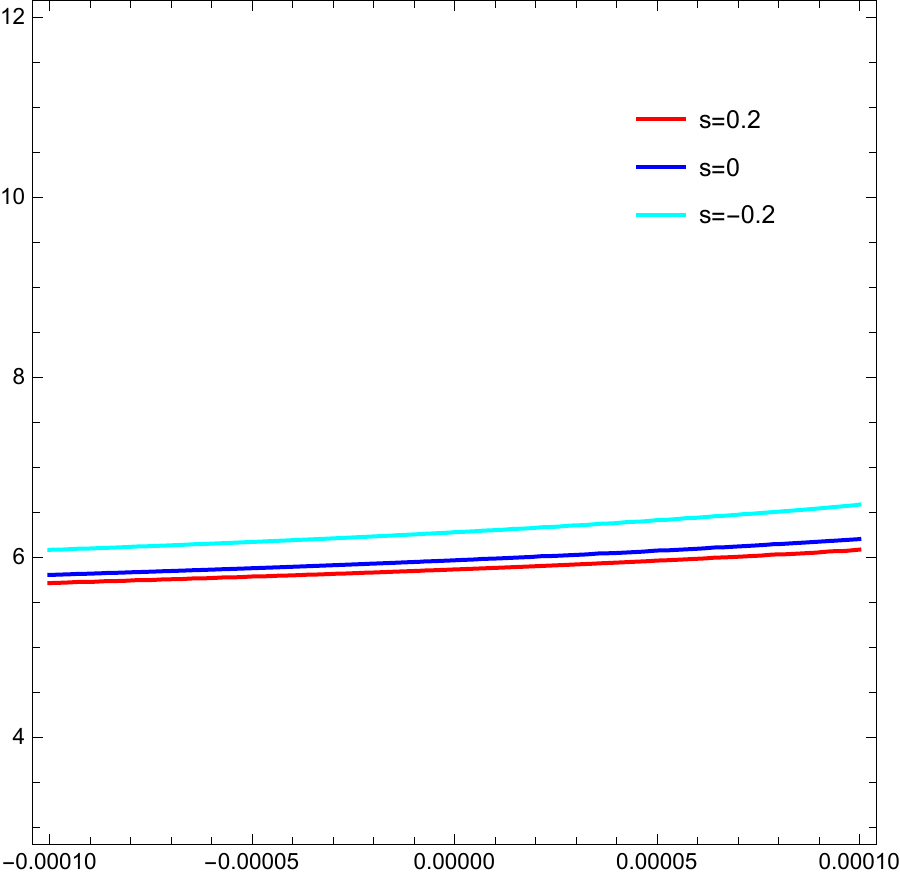}
	\caption{ISCO radius in term of $L$ for co-rotating test body, with $a/M=0.01$ .}\label{fig:ISCO-co4a=.1}
\end{figure}

As it is obvious from these plots, the rule for sign of the spin $s$ is different in co-rotating and counter-rotating cases for both slowly-rotating and extremal Kerr-(A)dS backgrounds. This is the same behavior that we saw in the $\mathcal{V}-r$ region, in which the Aschenbach effect occurs.{\bf{ We note that following \cite{Abramowicz:2004tm,Stuchlik:2006xa}, with the derived effective potential  Eq.\eqref{eff}, we are going to relate the current studies in the case of spinning test body moving around Kerr-(A)dS black hole to the QPOs phenomena observed in several black hole systems. However, we leave this studies to future publications.  
}}



\section*{References}
\begin{thebibliography}{10}


\bibitem{Abbott:2016blz}
B.~P.~Abbott \textit{et al.} [LIGO Scientific and Virgo],
Phys. Rev. Lett. \textbf{116}, no.6, 061102 (2016)
doi:10.1103/PhysRevLett.116.061102
[arXiv:1602.03837 [gr-qc]].


\bibitem{Eckart:2017bhq}
A.~Eckart, A.~H\"uttemann, C.~Kiefer, S.~Britzen, M.~Zaja\v{c}ek, C.~L\"ammerzahl, M.~St\"ockler, M.~Valencia-S, V.~Karas and M.~Garc\'\i{}a-Mar\'\i{}n,
Found. Phys. \textbf{47}, no.5, 553-624 (2017)
[arXiv:1703.09118 [astro-ph.HE]].


\bibitem{Condon:2018eqx}
J.~J.~Condon and A.~M.~Matthews,
Publ. Astron. Soc. Pac. \textbf{130}, no.989, 073001 (2018)
[arXiv:1804.10047 [astro-ph.CO]].



\bibitem{Guica:2008mu}
M.~Guica, T.~Hartman, W.~Song and A.~Strominger,
Phys. Rev. D \textbf{80}, 124008 (2009)
[arXiv:0809.4266 [hep-th]].


\bibitem{Abuter:2018drb}
R.~Abuter \textit{et al.} [GRAVITY],
Astron. Astrophys. \textbf{615}, L15 (2018)
[arXiv:1807.09409 [astro-ph.GA]].

\bibitem{Babak:2017tow}
S.~Babak, J.~Gair, A.~Sesana, E.~Barausse, C.~F.~Sopuerta, C.~P.~L.~Berry, E.~Berti, P.~Amaro-Seoane, A.~Petiteau and A.~Klein,
Phys. Rev. D \textbf{95}, no.10, 103012 (2017)
[arXiv:1703.09722 [gr-qc]].

\bibitem{McClintock:2011zq}
J.~E.~McClintock, R.~Narayan, S.~W.~Davis, L.~Gou, A.~Kulkarni, J.~A.~Orosz, R.~F.~Penna, R.~A.~Remillard and J.~F.~Steiner,
Class. Quant. Grav. \textbf{28}, 114009 (2011)
[arXiv:1101.0811 [astro-ph.HE]].

\bibitem{Laura}
B.~Laura,
"Measuring the Angular Momentum of Supermassive Black Holes",2013.

\bibitem{Aschenbach:2004kj}
B.~Aschenbach,
Astron. Astrophys. \textbf{425}, 1075-1082 (2004)
[arXiv:astro-ph/0406545 [astro-ph]].

\bibitem{Aschenbach:2006cj}
B.~Aschenbach,
Chin. J. Astron. Astrophys. Suppl.1, \textbf{6}, 01221 (2006)
[arXiv:astro-ph/0603193 [astro-ph]].

\bibitem[Gravity Collaboration \emph{et al.}(2018)]{2018A&A...618L..10G}Gravity Collaboration, Abuter, R., Amorim, A., Baub{\"o}ck, M., Berger, J.P., Bonnet, H., and, ...: 2018, {\it Astronomy and Astrophysics} {\bf 618}, L10.
[arXiv:1810.12641 [astro-ph]]



\bibitem[Tursunov \emph{et al.}(2020)]{2020ApJ...897...99T} A. Tursunov, M. Zaja{\v{c}}ek, A. Eckart, M. Kolo{\v{s}}, S. Britzen, Z. Stuchl{\'\i}k, B. Czerny and V. Karas,  2020, \apj, {\bf 897}, 1.


\bibitem[Thorne(1974)]{1974ApJ...191..507T} Thorne, K.~S.\ 1974, \apj, 191, 507. 

\bibitem{Stuchlik:2007sta}
Z.~Stuchl\'\i{}k, P.~{Slan{\'y}}, G.~{T{\"o}r{\"o}k} and M.~A.~Abramowicz, 
Astron. Astrophys. \textbf{463}, 3, 807 (2007)


\bibitem{Stuchlik:2006xa}
Z.~Stuchlik, P.~Slany and G.~Torok,
PoS \textbf{MQW6}, 095 (2006)
[arXiv:astro-ph/0612439 [astro-ph]].

\bibitem{Kluzniak:2001abr}
W.~Kluzniak  and M.~A.~Abramowicz, 
Acta Phys. Pol. B \textbf{32}, 3605 (2001) 

\bibitem{Abramowicz:2003bbk}
M.~A.~Abramowicz, T. Bulik, M. Bursa  and W.~Kluzniak, 
Astron. Astrophys. \textbf{404}, L21 (2003) 

\bibitem{Torok:2005aks} 
G.~{T{\"o}r{\"o}k}, M.~A.~Abramowicz, W.~Kluzniak and Z.~Stuchl\'\i{}k, ,
Astron. Astrophys. \textbf{436}, 1, (2005)

\bibitem{Kolos:2017ojf}
M.~Kolo\v{s}, A.~Tursunov and Z.~Stuchl\'\i{}k,
Eur. Phys. J. C  \textbf{77}, 12, 860 (2017) 
[arXiv:1707.02224 [astro-ph.HE]].

\bibitem{Tursunov:2018H}
A.~Tursunov and M.~Kolo\v{s}, 
Phys. Atom. Nuclei \textbf{81}, 279 (2018) 
[arXiv:1803.08144 [astro-ph.HE]].


\bibitem{Stuchlik:2004wk}
Z.~Stuchlik, P.~Slany, G.~Torok and M.~A.~Abramowicz,
Phys. Rev. D \textbf{71}, 024037 (2005)
[arXiv:gr-qc/0411091 [gr-qc]].

 
\bibitem{Stuchlik:2011sw}
Z.~Stuchl\'\i{}k, M.~Blaschke and P.~{Slan{\'y}},
Class. Quant. Grav. \textbf{28}, 175002 (2011)
[arXiv:1108.0191 [gr-qc]].


\bibitem{Tursunov:2016dss}
A.~Tursunov, Z.~Stuchl\'\i{}k and M.~Kolo\v{s},
Phys. Rev. D \textbf{93}, no.8, 084012 (2016)
[arXiv:1603.07264 [gr-qc]].


\bibitem{2020Univ....6...26S}
Z.~Stuchl\'\i{}k, M.~Kolo\v{s}, J.~{Kov{\'a}{\v{r}}}, P.~{Slan{\'y}} and A.~Tursunov, 
Universe \textbf{6}, no.2, 26 (2020)


\bibitem{Khodagholizadeh:2020sex}
J.~Khodagholizadeh, V.~Perlick and A.~Vahedi,
Phys. Rev. D \textbf{102}, no.2, 024021 (2020)
[arXiv:2002.04701 [gr-qc]].








\bibitem{Mathisson1937}
M.~Mathisson, Acta Phys. Pol. {\bf 6}, 163 (1937).

\bibitem{Papapetrou1951}
A.~Papapetrou, Proc. R. Soc. A {\bf 209}, 248 (1951).

\bibitem{Dixon1964}
W. G. Dixon, Nuovo Cim. {\bf 34}, 317 (1964).

\bibitem{Slany:2007gp}
P.~{Slan{\'y}} and Z.~Stuchl\'\i{}k,
[arXiv:0709.0803 [gr-qc]].
\bibitem{Mueller:2007tf}
A.~Mueller and B.~Aschenbach,
Class. Quant. Grav. \textbf{24}, 2637-2644 (2007)
doi:10.1088/0264-9381/24/10/009
[arXiv:0704.3963 [gr-qc]].

\bibitem{Semerak:1999qc}
O.~Semerak,
Mon. Not. Roy. Astron. Soc. \textbf{308}, 863-875 (1999)



\bibitem{Tulczyjew1959} W. Tulczyjew, 
Acta Phys. Pol. {\bf 18}, 393 (1959).

\bibitem{Dixon1970} W. G. Dixon, 
Proc. Roy. Soc. Lond. A {\bf 314}, 499 (1970).

\bibitem{Frenkel1926} J. Frenkel, 
Z. Phys. 37, 243 (1926), Nature {\bf 117}, 653 (1926).

\bibitem{Pirani1956} F. A. E. Pirani, 
Acta Phys. Pol. {\bf 15}, 389 (1956).

\bibitem{Chicone:2005jj}
C.~Chicone, B.~Mashhoon and B.~Punsly,
Phys. Lett. A \textbf{343}, 1-7 (2005)
[arXiv:gr-qc/0504146 [gr-qc]].

\bibitem{CARTER1968}
B.~ Carter,
"A new family of einstein spaces", Physics Letters A(1968).

\bibitem{Zhang:2018omr}
Y.~P.~Zhang, S.~W.~Wei, P.~Amaro-Seoane, J.~Yang and Y.~X.~Liu,
[arXiv:1812.06345 [gr-qc]].

\bibitem{Hackmann:2014tga}
E.~Hackmann, C.~L\"ammerzahl, Y.~N.~Obukhov, D.~Puetzfeld and I.~Schaffer,
Phys. Rev. D \textbf{90}, no.6, 064035 (2014)
[arXiv:1408.1773 [gr-qc]].

\bibitem{p2015}
Dirk Puetzfeld and  Claus L\"{a}mmerzahl and Bernard Schutz,
"Equations of Motion in Relativistic Gravity",(2015)


\bibitem{Ehlers1977}
J{\"{u}}rgen Ehlers and Ekkart Rudolph,
"Dynamics of extended bodies in general relativity center-of-mass description and quasirigidity",(1977).


\bibitem{Kerr:2003bp}
A.~W.~Kerr, J.~C.~Hauck and B.~Mashhoon,
Class. Quant. Grav. \textbf{20}, 2727 (2003)
[arXiv:gr-qc/0301057 [gr-qc]].

\bibitem{zamo}
J.~M.~Bardeen, W.~H.~Press and S.~A.~Teukolsky,
Astrophys.\ J.\  {\bf 178}, 347 (1972).

\bibitem{Stuchlik:2003dt}
Z.~Stuchl\'\i{}k and P.~{Slan{\'y}},
Phys. Rev. D \textbf{69}, 064001 (2004)
[arXiv:gr-qc/0307049 [gr-qc]].

\bibitem{Stuchlik:1983baic}
Z.~Stuchlik, 
Bull. Astron. Inst. Czech. \textbf{34}, 129 (1983)

\bibitem{Stuchlik:2004prdhle}
Z.~Stuchl\'\i{}k and J.~Hledik,
Phys. Rev. D \textbf{60}, 044006 (1999)



\bibitem{Abramowicz:2004tm}
M.~A.~Abramowicz and W.~Kluzniak,
Astrophys. Space Sci. \textbf{300}, 127 (2005)
[arXiv:astro-ph/0411709 [astro-ph]].


\bibitem{Kluzniak:2001bgr}
Kluzniak, W., Bulik, T., Gondek-Rosinska, D., 
Proceedings of the Fourth INTEGRAL Workshop, 4-8 September 2000,Alicante, Spain. ESA SP-459, Noordwijk, 
ESA Publications Division, 301 (2001) 



\bibitem{Saijo1998}
M.~Saijo, K.-i.~Maeda, M.~Shibata and Y.~Mino,
Phys. Rev. D {\bf 58}, 064005 (1998).

\bibitem{Abram79}
M.~A.~Abramowicz, M.~Calvani, Monthly Notices of the Royal Astronomical Society \textbf{189}, 3, 621 (1979).

\bibitem{Kerner:2001cw}
R.~Kerner, J.~W.~van Holten and R.~Colistete, Jr.,
Class. Quant. Grav. \textbf{18}, 4725-4742 (2001)
[arXiv:gr-qc/0102099 [gr-qc]].


\bibitem{Zhang:2018ocv}
S.~Zhang, Y.~Liu and X.~Zhang,
Phys. Rev. D \textbf{99}, no.6, 064022 (2019)
[arXiv:1812.10702 [gr-qc]].

\bibitem{Kraniotis:2005zm}
G.~V.~Kraniotis,
Class. Quant. Grav. \textbf{22}, 4391-4424 (2005)
[arXiv:gr-qc/0507056 [gr-qc]].


\bibitem{Frolov:2017kze}
V.~Frolov, P.~Krtous and D.~Kubiznak,
Living Rev. Rel. \textbf{20}, no.1, 6 (2017)
[arXiv:1705.05482 [gr-qc]].


\bibitem{Jefremov:2015gza}
P.~I.~Jefremov, O.~Y.~Tsupko and G.~S.~Bisnovatyi-Kogan,
Phys. Rev. D \textbf{91}, no.12, 124030 (2015)
[arXiv:1503.07060 [gr-qc]].



\bibitem{Howes}
Robert.~J.~.Howes
"Effects of a Positive Cosmological Constant
on Circular Orbits in the Reissner-Nordstrom, Schwarzschild, and Kerr Fields", Springer, 829–835 (1980). 


\end{thebibliography}
\end{document}